\documentclass[]{aa}
\usepackage{graphicx}
\usepackage{amsmath}
\usepackage{subcaption} 
\usepackage{float}
\usepackage{placeins}

\usepackage{natbib}
\bibpunct{(}{)}{;}{a}{}{,} 

\usepackage[colorlinks,allcolors=blue]{hyperref}

\def\H0{{\rm ~km~s^{-1}~Mpc^{-1}}}

\begin{document} 

\title{An outflow in the Seyfert ESO 362-G18 revealed by Gemini-GMOS/IFU observations}

          \author{Humire, Pedro K., 
          \inst{1}
          Nagar, Neil M.,
          \inst{1}
		  Finlez, Carolina,
          \inst{1}
          Firpo, Verónica,
          \inst{2,3}
          Slater, Roy,
          \inst{1}
          Lena, Davide,
          \inst{4,5}          
          Soto, Pamela R.,
          \inst{1}
          Muñoz, Dania,
          \inst{1}
          Riffel, Rogemar A.,
          \inst{6}
          Schmitt, H.R.,
          \inst{7}
          Kraemer, S.B.,
          \inst{8}
          Schnorr-M\"uller, Allan,
          \inst{9}
          Fischer, T.C.,
          \inst{10}
          Robinson, Andrew,
          \inst{11}
          Storchi-Bergmann, Thaisa,
          \inst{9}
          Crenshaw, Mike
          \inst{12}
          \and
          Elvis, Martin S.
          \inst{13}
         }


\authorrunning{P. Humire et al.}
\titlerunning{ESO362-G18 with Gemini-GMOS/IFU}

          \institute{Departamento de Astronom\'ia, Universidad de Concepci\'on, Casilla 160-C, Concepci\'on, Chile\\
              \email{phumire@astro-udec.cl}
          \and
               Departamento de Física y Astronomía, Universidad de La Serena, La Serena, Chile 
          \and
               Gemini Observatory, Southern Operations centre, La Serena, Chile
          \and 
               SRON, Netherlands Institute for Space Research, Sorbonnelaan 2, NL\-3584 CA Utrecht, the Netherlands
           \and    
               Department of Astrophysics/IMAPP, Radboud University, Nijmegen, PO Box 9010, NL-6500 GL Nijmegen, the Netherlands
          \and
               Universidade Federal de Santa Maria, Departamento de F\'isica, Centro de Ciencias Naturais e Exatas, 97105-900, Santa Maria, RS, Brazil  
          \and  
               Naval Research Laboratory, Washington, DC 20375, USA
          \and
               Institute for Astrophysics and Computational Sciences, Department of Physics, The Catholic University of America, Washington, DC 20064, USA
          \and
               Instituto de F\'isica, CP 15015,Universidade Federal do Rio Grande do Sul, 91501-970, Porto Alegre, RS, Brazil
          \and 
               Astrophysics Science Division, Goddard Space Flight centre, Code 665, Greenbelt, MD 20771, USA
          \and 
               Department of Physics, Rochester Institute of Technology, 84 Lomb Memorial Drive, Rochester, NY 14623, USA
          \and 
               Department of Physics and Astronomy, Georgia State University, Astronomy Offices, 25 Park Place, Suite 605, Atlanta, GA 30303, USA
          \and     
               Harvard-Smithsonian centre for Astrophysics
60 Garden St., ms.6, Cambridge MA 02138 USA
               }


   \date{Received xxx, 2017; accepted xxx, 2017}

\abstract{We present two-dimensional stellar and gaseous kinematics of the inner 0.7 $\times$ 1.2 kpc$^{2}$ of the Seyfert 1.5 galaxy ESO 362-G18, derived from optical (4092-7338 \AA) spectra obtained with the GMOS integral field spectrograph on the Gemini South telescope at a spatial resolution of $\approx$170 pc and spectral resolution of 36 km s$^{-1}$. ESO 362-G18 is a strongly perturbed galaxy of morphological type Sa or S0/a, with a minor merger approaching along the NE direction. Previous studies have shown that the [O\,{\sc iii}] emission shows a fan-shaped extension of $\approx$ 10\arcsec\ to the SE. We detect the [O\,{\sc iii}] doublet, [N\,{\sc ii}] and H${\alpha}$ emission lines throughout our FOV. 

The stellar kinematics is dominated by circular motions in the galaxy plane, with a kinematic position angle of $\approx$137$^{\circ}$ and is centred approximately on the continuum peak. The gas kinematics is also dominated by rotation, with kinematic position angles ranging from 122$^{\circ}$ to 139$^{\circ}$, projected velocity amplitudes of the order of 100 km s$^{-1}$, and a mean velocity dispersion of 100 km s$^{-1}$. 
A double-Gaussian fit to the [O\,{\sc iii}]$\lambda$5007 and H${\alpha}$ lines, which have the highest signal to noise ratios of the emission lines, reveal two kinematic components: 
(1) a component at lower radial velocities which we interpret as gas rotating in the galactic disk; and 
(2) a component with line of sight (LOS) velocities 100--250 km s$^{-1}$ higher than the systemic velocity, interpreted as originating in the outflowing gas within the AGN ionization cone.  

We estimate a mass outflow rate of 7.4 $\times$ 10$^{-2}$ M$_{\odot}$ yr$^{-1}$ in the SE ionization cone (this rate doubles if we assume a biconical configuration), and a mass accretion rate on the supermassive black hole (SMBH) of 2.2 $\times$ 10$^{-2}$ M$_{\odot}$ yr$^{-1}$. The total ionized gas mass within $\sim$84 pc of the nucleus is 3.3 $\times$ 10$^{5}$ M$_{\odot}$; infall
velocities of $\sim$34 km s$^{-1}$ in this gas would be required to feed both the outflow and SMBH accretion.
}


   \maketitle
   
%

\section{Introduction}

It is now widely accepted that the intense radiation emitted by an active galactic nucleus (AGN) is due to accretion onto a supermassive black hole (SMBH) \citep{Lynden-Bell1969, Begelman1984} in the mass range $\sim$ 10$^{6}$-10$^{9}\,{\rm M_\odot}$. However, the mechanisms responsible for transferring the mass from galactic (kpc) scales down to nuclear scales (sub-parsec) to feed the SMBH are still under debate. This has been the subject of many theoretical and observational studies \citep{Shlosman1990,Maciejewski2004a,Maciejewski2004b,Knapen2005,Emsellem2006,SchnorrMuller2014a,SchnorrMuller2014b,SchnorrMuller2017}. 


Theoretical studies and simulations have shown that non-axisymmetric potentials efficiently promote gas inflow towards the inner regions of galaxies \citep{Englmaier2004}. Close encounters and galactic mergers have been identified as a mechanism capable of driving gas from tens of kiloparsecs down to a few kiloparsecs \citep{Hernquist1989, DiMatteo2005}. Major mergers are apparently a requirement for triggering the most luminous AGNs \citep{Treister2012}. Simulations by \citet{Hopkins2010} suggest that in gas-rich systems, at scales of 10 to 100 pc, inflows are achieved through a system of gravitational instabilities over a wide range of morphologies such as nuclear spirals, bars, rings, barred rings, clumpy disks, and streams.

Indeed, several observations support the hypothesis that large-scale bars channel the gas to the centres of galaxies \citep{Crenshaw2003a}. Recent studies have concluded that there is an excess of bars among Seyfert galaxies as compared to non-active galaxies of about 75\% versus 57\%, respectively \citep{Knapen2000, Laine2002, Laurikainen2004}. Further, structures such as disks or small-scale nuclear bars and the associated spiral arms are often found in the inner kiloparsec of active galaxies \citep{Erwin1999, Pogge2002, Laine2003, Combes2014}. In general, the most common nuclear structures are dusty spirals, estimated to reside in more than half of active and inactive galaxies \citep[71\% and 61\%, respectively;][]{Martini2003}. 

\citet{Simoes2007} reported a marked difference in the dust and gas content of early-type active and non-active galaxies: the former always have dusty structures and only 25\% of the latter have such structures. Thus, a reservoir of gas and dust is required for the nuclear activity suggesting that the dusty structures are tracers of feeding channels to the AGN. This fact, along with the enhanced frequency of dusty spirals, supports the hypothesis that nuclear spirals are a mechanism for fueling the SMBH, transporting the gas from kiloparsec scales down to a few tens of parsecs of the nucleus.


Accretion onto the SMBH requires the removal of angular momentum, which can be achieved not only through gravitational torques, but also via outflows or winds \citep{Bridle1984}. The most powerful of these outflows are produced by the interaction between the ionized gas and magnetic field \citep{BisnovatyiKogan2001} reaching velocities of up to 1000 km s$^{-1}$ \citep{Rupke2011, Greene2012} and outflow rates several times larger than host galaxy star formation rates (hereafter SFR) \citep{Sturm2011}. Massive AGN-driven outflows have been observed in many AGN, from Seyfert galaxies to quasars at low \citep{Morganti2007} and high redshifts \citep{Nesvadba2011}, and could dramatically affect the evolution of galaxies due to the large amounts of energy they feed back into the interstellar medium \citep{DiMatteo2005}. At the less powerful end, studies of nearby Seyferts show that compact outflows ($\sim$100 pc in extent) with velocities of $\sim$100 km s$^{-1}$ and mass outflow rates of a few solar masses per year are common even in low-luminosity AGNs \citep[e.g.][]{MullerSanchez2011, Davies2014}.

At low outflow velocities, it can be difficult to identify if AGNs or host galaxy starbursts are responsible for the outflow: a cut-off of 500 km s$^{-1}$ is often used \citep{Fabian2012,Cicone2014} to differentiate the two. Identifying low-velocity outflows requires relatively high spectral resolutions and two-dimensional spectroscopy (integral-field spectrographs) to disentangle the different velocity components present: from the galactic disk and from outflow(s) and/or inflow(s) \citep{Storchi-Bergmann2010,SchnorrMuller2014a}. Moreover, in some special cases these outflows are detected more frequently as redshifted, rather than blueshifted, winds since the light from the ionized regions reaches us preferentially from the receding side of the outflow which, for our LOS, is more illuminated by the AGN \cite[e.g.][]{Lena2015}, and 
the kinematics can often be modelled as a combination of a biconical outflow and a rotating disk coincident with the molecular gas \citep{MullerSanchez2011, Fischer2013}. A better understanding of low-velocity outflows in nearby Seyferts is important to understand the kinematics of Seyfert galaxies at higher redshift, which could share the same model.

Outflows could be very important for the evolution of galaxies because they can be the most efficient way for the interaction between the AGN and its host galaxy, a process called AGN feedback, affecting the interstellar medium and star formation. Several works explored whether this process triggers (positive feedback) or extinguishes (negative feedback) the host galaxy star formation \citep{Fabian2012, Karouzos2014}. Empirical scaling relations between the masses of the SMBH and the host-galaxy bulge \citep[e.g.][]{Gultekin2009}, and between the AGN luminosity and the molecular outflow velocity \citep{Sturm2011} or dynamical mass \citep{Lin2016}, have motivated a more intensive study of outflows.

In this work, we present results obtained from integral field spectroscopy observations of the nuclear region of ESO 362-G18 (a.k.a. MCG 05-13-17), a nearby galaxy of morphological type Sa \citep*[][hereafter MGT]{Malkan1998} or S0/a \citep[RC3;][]{deVaucouleurs1991} harbouring a Seyfert 1.5 nucleus \citep{Bennert2006}. ESO 362-G18 has a redshift of 0.012445 and a systemic velocity of 3731 km s$^{-1}$ \citep{Paturel2003} or 3707 km s$^{-1}$ \citep{Makarov2014}; we consider the former estimate since it represents our data very well. Assuming H$_{0}$=73.0${{\rm ~km~s^{-1}~Mpc^{-1}}}$, this corresponds to a distance of 50.8 Mpc and a linear scale of 246 pc arcsec$^{-1}$. Previous studies estimated morphological position angles (PA) between 110$^{\circ}$ and 160$^{\circ}$ \citep[RC3,][and references therein]{Fraquelli2000} and a disk inclination (i), ranging from 37$^{\circ}$ to 54$^{\circ}$ \citep[RC3, respectively]{Fraquelli2000}. ESO 362-G18 has been studied in the radio, near-infrared (NIR), optical, UV, and X-ray; its nucleus has typically been classified as Seyfert 1  \citep[MGT,][]{Mulchaey1996,RodriguezArdila2000,Fraquelli2000,AgisGonzalez2014}.

Previous studies indicate that ESO 362-G18 is a highly disturbed galaxy \citep{Mulchaey1996} with a ``long faint plume'' to the NE \citep{Corwin1985}; this plume is likely an infalling less massive galaxy 10\arcsec\ to the NE, i.e. a minor merger. The emission-line maps of \citet{Mulchaey1996} revealed strong [O\,{\sc iii}] emission centred near the continuum peak with a fan-shaped emission of $\sim$10\arcsec\ in the SE direction, roughly along the host galaxy major axis and coincident with the strongest H$\alpha$ emission that is more symmetrically distributed about the nucleus. \citet{Mulchaey1996} estimated that the highest excitation gas is located $\sim$7\arcsec\ SE from the nucleus on one edge of the ionization cone, but \citet{Bennert2006} found that only the central $\pm$3\arcsec\ show line ratios typical of AGN ionized gas, and confirm the suggestion of \citet{Fraquelli2000} that the ionization parameter is peaked in the nucleus and rapidly decreases within the narrow line region (NLR) based on the increased [OII]/[O\,{\sc iii}] ratio. \citet{Fraquelli2000} also suggested that the nuclear continuum ionizes the gas in the disk along PA = 158$^{\circ}$, giving rise to the fan-shaped region observed in [O\,{\sc iii}]. Arcsecond resolution centimeter radio maps of ESO 362-G18 do not show any obvious extensions \citep{Nagar1999}. \citet{Bennert2006} find that the spectra out to r $\sim$ 11\arcsec\ NW and out to r $\sim$ 6\arcsec\ SE have line ratios that fall in the regime of H\,{\sc ii} regions. Indeed, \citet{Tsvetanov1995} identified 38 H\,{\sc ii} regions in their ground-based H$\alpha$+[N\,{\sc ii}] image of  ESO 362-G18, distributed in a cloud around the nucleus with distances between 3-18\arcsec. 

The nuclear optical spectrum is dominated by broad permitted lines and narrow permitted and forbidden lines \citep{RodriguezArdila2000} and shows a featureless nuclear continuum due to the AGN \citep{Fraquelli2000}, and main stellar features of Ca II K, G band, Mg I b and Na I D, as well as high order Balmer absorptions lines outside the nucleus. A broad Balmer decrement H$\alpha$$_{broad}$/H$\beta$$_{broad}$ of 5.7 indicates a slightly higher reddening of the broad line region (BLR) with respect to the central NLR \citep{Bennert2006}. 

Near-infrared spectroscopy \citep{Riffel2006} shows strong, broad H I, and He I lines with a full width at half maximum (FWHM) of $\approx$4500 km s$^{-1}$ and $\approx$5400 km s$^{-1}$, respectively. Besides, numerous forbidden lines are seen, including high-ionization lines of [S\,{\sc ix}], [Si\,{\sc x}], and [Si\,{\sc vi}]. The Br$\gamma$ and $H_{2}$ molecular emission lines are observed as well, although they are intrinsically weak. The NIR continuum emission present stellar absorption features of Ca II, CO in $H$ band, and the 2.3 $\mu$m CO band heads in $K$ band on top of a steep power-law like the continuum.

The most recent detailed study of this galaxy has been conducted by \citet{AgisGonzalez2014}. Their most important result was to detect a large variability in X-ray absorption which they explain as a clumpy, dusty torus lying in a compact region within $\sim$ 50r$_{g}$ (probably with 7 r$_{g}$, 1r$_{g}$ = GM$_{BH}$/c$^{2}$) from the central black hole. They also estimated an inner accretion disk inclination of i=53$^{\circ}$ $\pm$ 5$^{\circ}$, i.e. aligned with the large-scale galaxy disk (RC3; 54$^{\circ}$).

This paper is organized as follows. Section 2 describes the observations, data processing, and analysis; Section 3 presents our results; Section 4 discusses the results and presents estimates of the mass outflow and inflow rates; and Section 5 presents our conclusions.


\section{Observations, data processing and analysis software}

The observations were obtained with the integral field unit of the Gemini Multi-Object Spectrograph \citep[GMOS-IFU,][]{Gemini_South} at the Gemini South telescope on the night of December 23, 2014 (Gemini project GS-2014B-Q-20). The observations were made in the one-slit mode of GMOS-IFU, in which the science IFU has a field of view (hereafter FOV) of 3\farcs5 $\times$ 5\arcsec. Two pointings, shifted by 0\farcs5 were observed so that the total sky coverage was 4\arcsec\ $\times$ 5\arcsec, centred on the nucleus. Two exposures of 900 seconds were made at each pointing with a shift in the central wavelength of 50\AA\ between the two. The seeing at the time of the science observations was  0\farcs7, as listed in the Gemini observations log, and we confirmed this value by fitting the luminosity profile of the broad line component of H$\alpha$ (see Sect. 3.5). This corresponds to a linear resolution of 172 pc at the distance of the galaxy. The spectroscopic standard star LTT\,1788 (V\,=\,13.16) was observed in a 360~s exposure $\sim$1 hr before observing ESO 362-G18, under similar atmospheric conditions and with the same instrument set-up.

The selected wavelength range was 4092-7338 \AA\, to cover the H$\beta\,\lambda$\,4861, [O\,{\sc iii}]$\,\lambda \lambda$\,4959,5007, H$\alpha$+[N\,{\sc ii}]\,$\lambda \lambda$\,6548,6583 and [S\,{\sc ii}]$\,\lambda \lambda$\,6716,6731 emission lines, observed with the grating GMOS B600-G5323 (set to central wavelength of either $\lambda$5700\,\AA{} or $\lambda$5750\,\AA{}) at a spectral resolution of R\,$\approx$\,3534 at $\lambda$6440\,\AA{} corresponding to an instrumental dispersion ($\sigma_{inst}$) of $\approx$ 36\,km\,s$^{-1}$. Wavelength calibration is expected to be accurate to the order of 8\,km\,s$^{-1}$.

The data reduction was performed using specific tasks developed for GMOS data in the GEMINI.GMOS version 1.13 package and generic tasks in IRAF\footnote{IRAF is distributed by the National Optical Astronomy Observatories, which are operated by de Association of Universities for Research in Astronomy, Inc., under cooperative agreement with the National Science Foundation.}. The reduction process \citep[see][]{Lena2014} comprised bias subtraction, flat-fielding, trimming, wavelength calibration, sky subtraction, relative flux calibration, building of the data cubes at a sampling of 0\farcs08$\times$0\farcs08, and finally the alignment and combination of the four data cubes. Owing to signal-to-noise (S/N) limitations, we used only the overlapping area of the two spatial pointings and also eliminated spaxels at the edge of the IFU. The final science FOV was thus 2\farcs8 $\times$ 4\farcs8. Sky subtraction, performed using spectra from the sky IFU. In Fig. 1, we note a residual telluric absorption at $\sim$6870\AA\, but the [S\,{\sc ii}]$\,\lambda \lambda$\,6716,6731 emission lines are not affected by these telluric absorptions.


Flux calibration was performed using the spectroscopic standard star LTT\,1788 (V\,=\,13.16) for which fluxes are tabulated every 50\,\AA. 


In order to measure the stellar kinematics and create an emission-line-only cube, we employed the penalized pixel fitting technique (pPXF) \citep{Capellari2004}, using single stellar population (SSP) templates derived within the MILES Stellar Library \citep{Sanchez-Blazquez2006}. These templates have  a spectral resolution of 2.51\,\text{\AA} (FWHM) or $\sigma_{inst}$ $\sim$58\,km\,s$^{-1}$ and cover a spectral range of 3525\text{\AA} to 7500\text{\AA}. 
 Although the spectral resolution of the MILES templates is lower than that of our science data (36\,km\,s$^{-1}$), the MILES Stellar Library give better results than, for example, the Indo-U.S. Library \citep[$\sigma_{inst}$ $\approx$ 30\,km\,s$^{-1}$,][]{Valdes2004}, since the later does not give optimal fits especially  within the inner seeing disk. Comparing the pPXF results obtained using the two libraries individually, we observe that the differences in the fits are within the errors for the most part. When running pPXF with the MILES templates we did not convolve either to a lower spectral resolution. This is valid as the intrinsic stellar velocity dispersion in each spaxel is almost always above 60\,km\,s$^{-1}$ (as confirmed when using the INDO-US template library). The resulting stellar velocity dispersion map was corrected for the instrumental resolution of the science data (36\,km\,s$^{-1}$). 

Spatial averages (over large and small apertures) of spectra over various regions of the cube were first used to identify the 20 template spectra most used in the fits. These 20 spectra were then used to fit all individual spaxels in the cube. Before running pPXF, we masked spectral regions covering all broad and narrow emission lines; note that the former are present mainly within the inner seeing disk of 0\farcs7 (172\,pc). We used a tenth order additive polynomial in pPXF to take away the effects of the continuum shape of the stellar templates, host galaxy, and any AGN power-law continuum. The resulting best-fit templates were used to create an emission-line-only spectrum for each spaxel. Examples of this process are shown in Fig. 2. 

The centroid velocities, velocity dispersions and the emission-line fluxes of the gas were initially obtained from the emission-line-only cube by fitting a single Gaussian to the H${\alpha}$, H${\beta}$, [N\,{\sc ii}], [O\,{\sc i}], [O\,{\sc iii}], [S\,{\sc ii}]$\,\lambda$\,6716 and [S\,{\sc ii}]$\,\lambda$\,6731 emission lines using FLUXER\footnote{Interactive IDL routine written by Christof Iserlohe. http://www.ciserlohe.de/fluxer/fluxer.html}, which allows us to determine the residual continuum level around the emission lines in an interactive way; this is necessary for the [S\,{\sc ii}] lines since they are very close to the broad component of H${\alpha}$. The resulting gaseous velocities are similar to those obtained from the Gas AND Absorption Line Fitting code \citep[GANDALF;][]{Sarzi2006} and PROFIT \citep{Riffel2010}. To obtain the final velocity and dispersion maps we performed a sigma clip of 3$\sigma$ for all radial and velocity dispersion maps except in the nuclear regions of [O\,{\sc iii}] (because of their high S/N ratio). 
We also performed a double-Gaussian fit to the [O\,{\sc iii}] and H${\alpha}$ emission lines using a series of Python codes. 
To decide whether the observed line profile is better fit with a single or double Gaussian, we used the corrected Akaike information criterion \citep{Akaike1974} with the additional caveats that all Gaussian amplitudes are positive.
All emission-line velocity dispersion maps were corrected for the instrumental resolution (36\,km\,s$^{-1}$).

  \begin{figure*}[!ht]
   \centering
   \includegraphics[width=\hsize, trim={0 1cm 0 0}]{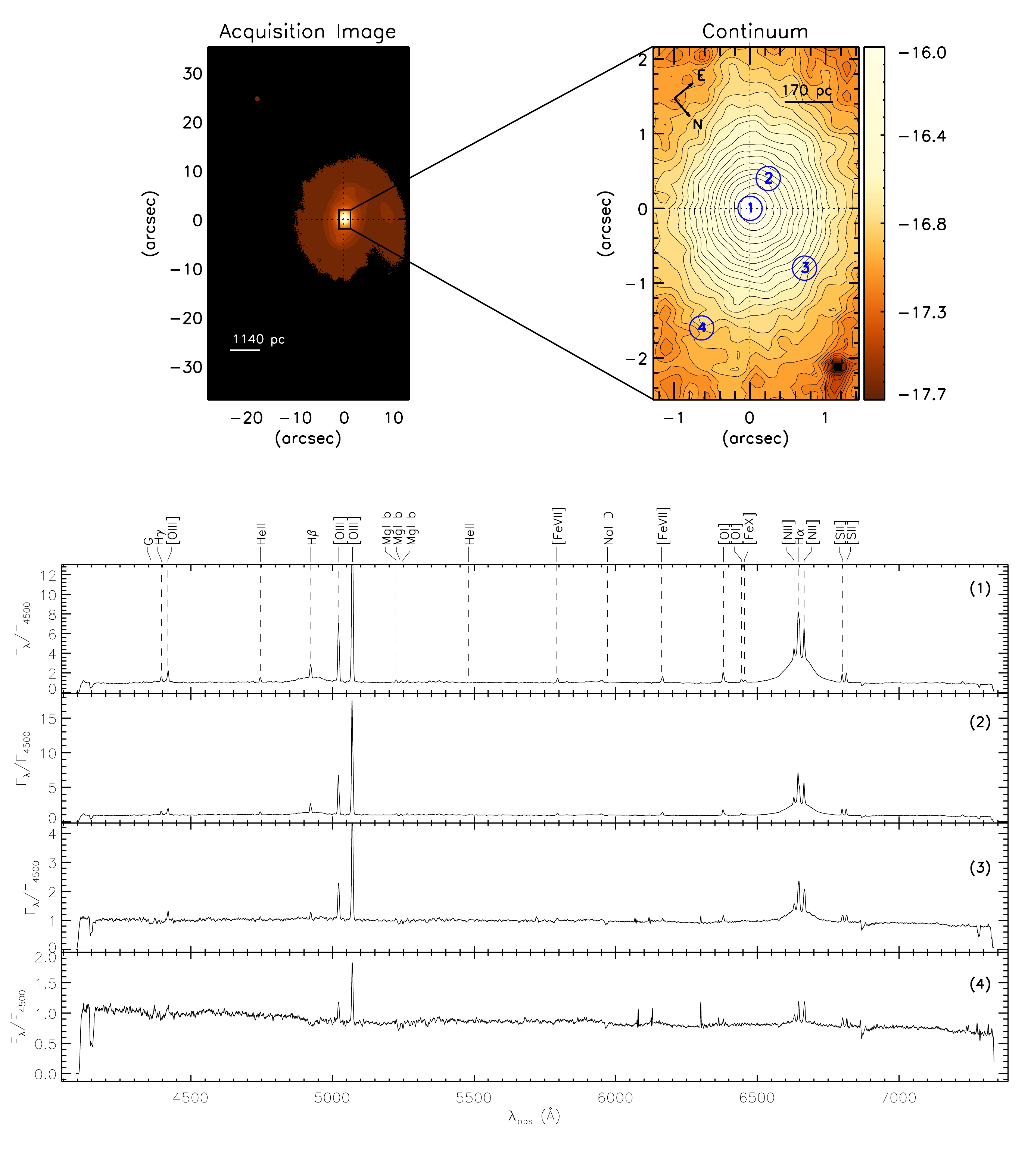}
      \caption{Top left: Acquisition Gemini GMOS image. The rectangle shows the FOV of the IFU observation. Top right: Continuum image of ESO\,362-G18 (see Section 3) is shown. Bottom: Observed-frame spectra corresponding to the regions are indicated as 1, 2, 3, and 4 in the IFU image.}
         \label{FigVibStab1}
   \end{figure*}

The kinematic PAs for both stars and gas were estimated using the code Kinemetry: a generalization of photometry to the higher moments of the LOS velocity distribution \citep{Krajnovic2006}. The systemic velocity (hereafter $V_{sys}$) of 3731\,km\,s$^{-1}$ was taken from \citet{Paturel2003}.

\begin{figure*}[!ht]
   \centering
   \includegraphics[width=0.7\textwidth, angle=90]{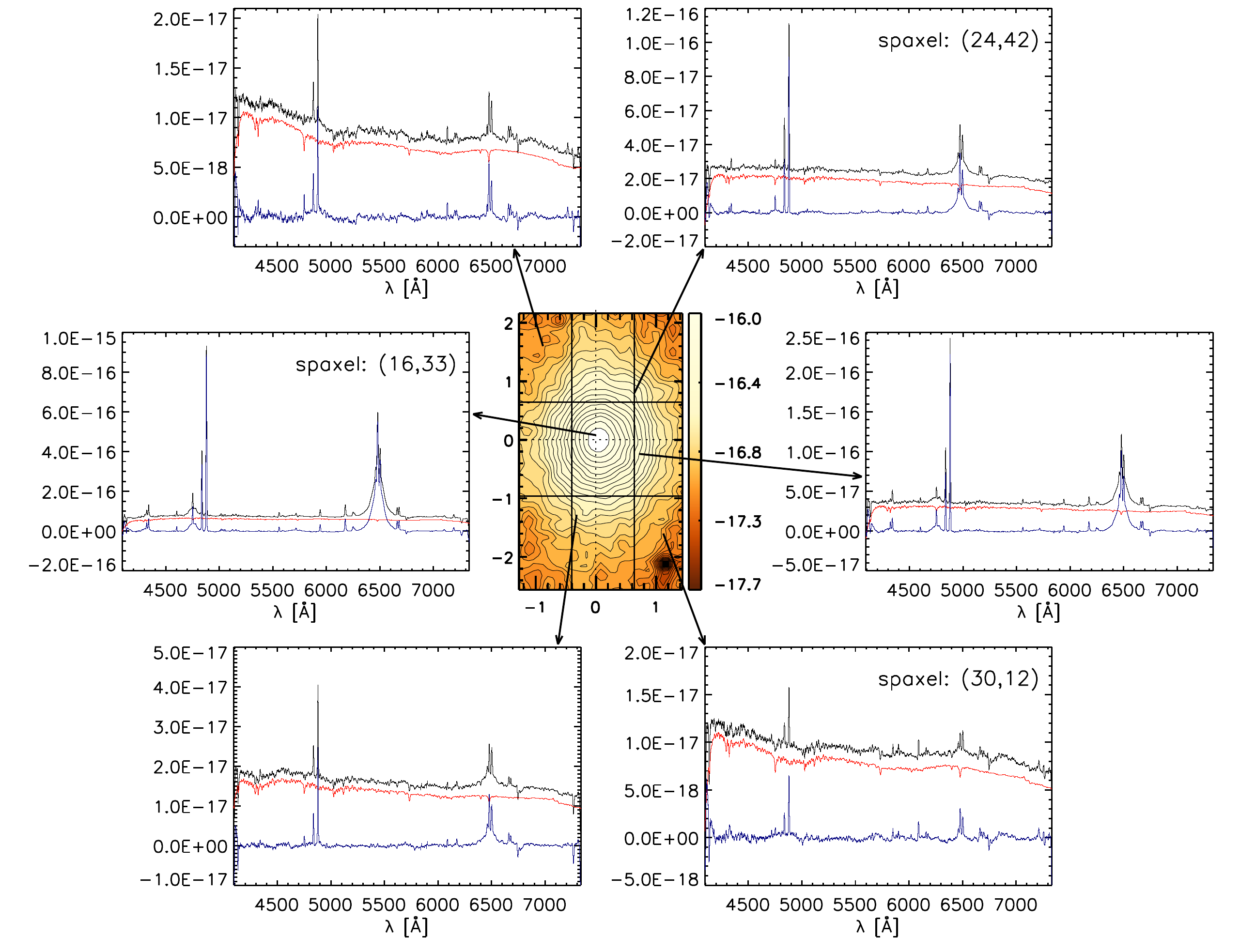}
      \caption{Fits of the Gemini GMOS spectra of ESO 362-G18 using the MILES Stellar Library \citep{Sanchez-Blazquez2006} for six spaxels, indicated by the arrows on the continuum image of ESO 362-G18 (the same as Fig. 1 top right panel). The spectra (black lines) and their corresponding fits (red lines) at various locations in the Gemini GMOS field are shown. The vertical and horizontal lines in this continuum image delimit the small apertures where spatial averages of spectra were taken to obtain the spectra most used by pPXF (whose contributions were more important for the best fit to the continuum). The vertical position of the best-fitting spectra obtained with pPXF \citep{Capellari2004} are slightly shifted for legibility; the residual (emission line only) spectra are presented on the bottom (blue lines). Axes units of the continuum image are arcseconds and the $y$ axes of the spectra have units of erg cm$^{-2}$ s$^{-1}$ \text{\AA}$^{-1}$.}
         \label{FigVibStab2}
   \end{figure*}

\section{Results}

The top left panel of Fig. 1 presents the Gemini acquisition image filter r of ESO\,362-G18, where the posited minor merger approaching from the NE direction is also clearly seen; the rectangle shows the FOV of the IFU. The top right panel shows the stellar continuum image obtained from our IFU data cube by integrating the flux within a spectral window from $\lambda$5345 \text{\AA} to $\lambda$ 5455 \text{\AA}. We assume a nuclear position that coincides with the location of the continuum peak in this image. 

In the bottom panel, we present four spectra from the locations indicated as 1 (nucleus), 2 and 3 (intermediate regions) and 4 (boundary region) in the IFU image and extracted within apertures of 0\farcs2 $\times$ 0\farcs2.
  The nuclear spectrum (identified as 1 in Fig. 1) shows broad H$\alpha$ and H${\beta}$ components, which led to the classification of ESO 362-G18 as a Seyfert 1 galaxy \citep{Fraquelli2000}, and also narrow [O\,{\sc iii}]\,$\lambda \lambda$\,4959,5007\,\text{\AA}, [O\,{\sc i}]\,$\lambda \lambda$ 6300,6363\,\text{\AA}, [N\,{\sc ii}]\,$\lambda \lambda$\,6548,6583 \text{\AA} and [S\,{\sc ii}]\,$\lambda \lambda$\,6717,6731 \text{\AA} emission lines. Large variations in the broad H${\beta}$ emission line of ESO 362-G18 has occasionally led to its classification as a Seyfert 1.5; such variations are not uncommon in Seyfert galaxies.  

We created a structure map of ESO 362-G18 (right panel of Fig. 5) by running the IDL routine unsharp\_mask.pro on an image obtained with WFPC3 (Wide Field Planetary Camera 3) through the filter F547M aboard the Hubble Space Telescope (hereafter HST; Program ID 13816). Inspection of the dust structure shows signs of spiral arms together with stronger obscuration to the SW than to the NE. We thus conclude that the SW is the near side of the galaxy. This is also consistent with flux asymmetries (Section 3.1) and a trailing spiral pattern (Section 3.2).

\begin{figure*}[!ht]
\centering
\includegraphics[width=\textwidth, trim={0.2cm 15cm -0.5cm 2cm},clip]{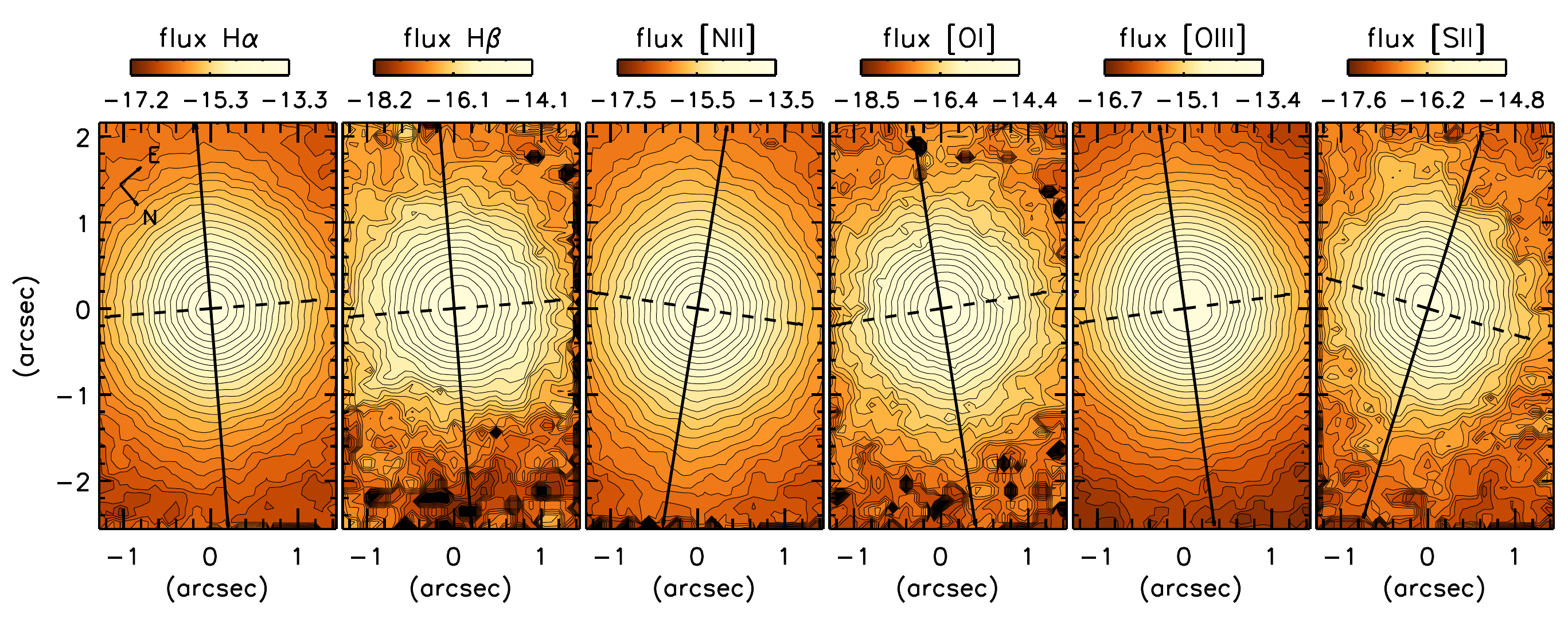}
  \caption{Maps of the H${\alpha}$, H${\beta}$, [N\,{\sc ii}], [O\,{\sc i}], [O\,{\sc iii}] and [S\,{\sc ii}] integrated fluxes in logarithmic scale (erg\,cm$^{-2}$\,s$^{-1}$ per pixel) following the colour bar above each panel. Axes are in arcseconds with respect to the nuclear continuum peak. The solid (dashed) lines indicate the major (minor) axis of the kinematics of the respective line, and the intersection between the lines indicate the continuum peak.}
\end{figure*}

\begin{figure*}[!ht]
\centering
    \includegraphics[width=\textwidth,trim={0 19cm 0 0}]{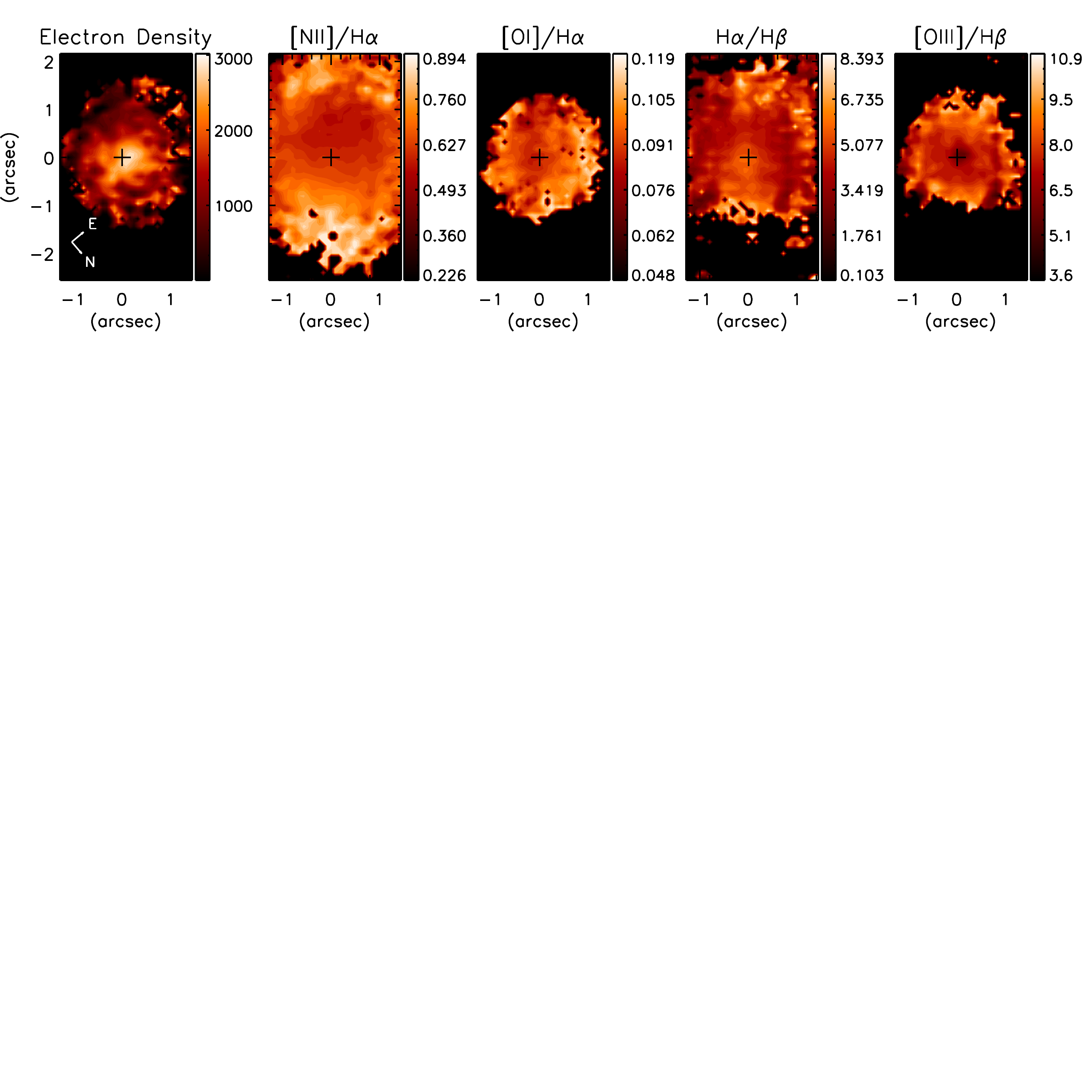}
  \caption{Maps of the estimated electron density (cm$^{-3}$) distribution and the line ratios [N\,{\sc ii}]/H$\alpha$, [O\,{\sc i}]/H$\alpha$, H$\alpha$/H$\beta$, and [O\,{\sc iii}]/H$\beta$. The cross represents the continuum peak and each panel follows the colour bar shown on its right.}
  \end{figure*} 
  
\subsection{\textbf{Morphology and excitation of the emitting gas}}

In Fig 3. we present the flux distributions derived from single Gaussian fits to the H$\alpha$, H$\beta$, [N\,{\sc ii}]$\,\lambda$6583\,\text{\AA}, [O\,{\sc i}]\,$\lambda$6300\,\text{\AA}, [O\,{\sc iii}]\,$\lambda$5007\,\text{\AA} and [S\,{\sc ii}]$\,\lambda$\,6716\,\text{\AA} emission lines. The flux distributions show a relatively smooth and symmetric pattern for all the emission lines, slightly elongated along the kinematic major axis (see Figures 6 and 7), as expected given the inclination. The highest fluxes are within the inner 1\arcsec\ (246\,pc). If the best-fitting two-dimensional Gaussian is subtracted from these flux distributions, the greatest asymmetries are found in the SW, implying a stronger presence of dust here. This supports our previous interpretation of the SW as the near side of the galaxy.   

Maps of the estimated electron density and the [N\,{\sc ii}]/H$\alpha$, [O\,{\sc i}]/H$\alpha$, H$\alpha$/H$\beta$, [O\,{\sc iii}]/H$\beta$ line ratios are presented in Fig. 4. The electron density was obtained from the [S\,{\sc ii}]$\,\lambda\lambda$\,6716/6731\,\text{\AA} line ratio assuming an electronic temperature of 10000 K \citep{Osterbrock2006}. The electron density reaches a peak value of $\sim$ 2900\,cm$^{-3}$ at the nucleus, decreasing to 1000 cm$^{-3}$ at 1\arcsec\ and 800\,cm$^{-3}$ at 1.5\arcsec\ from the nucleus. These values are in agreement with those obtained by \citet{Bennert2006} who estimated values from 1000\,cm$^{-3}$ up to 2500\,cm$^{-3}$ in that region.



The [N\,{\sc ii}]/H$\alpha$ line ratio shows values of 0.55-0.78 within the inner 1\arcsec\, and reach its highest values of close to 0.9 in a nuclear-centred ring of radius 1.5.\arcsec\ The [O\,{\sc iii}]/H$\beta$ ratio varies between 5.6 and 9 in the inner 0.5\arcsec\  and has a depression in the nucleus and increasing to 11 at 1.\arcsec\ Taking into account both the line ratios, the values can be considered typical of Seyfert galaxies \citep{CidFernandes2010}.

\subsection{\textbf{Stellar kinematics}}

The stellar velocity (V$_{\star}$) field, obtained from pPXF, is shown in the left panel of Fig. 5. This field displays a rotation pattern reaching amplitudes of $\approx$ 75\,km\,s$^{-1}$ within our FOV; the line of nodes are orientated approximately along the NW–SE direction with the SE side approaching and the NW side receding. With our adopted orientation, this implies trailing spiral arms as expected. The stellar velocity dispersion (fourth panel of Fig. 5) reaches values of 120\,km\,s$^{-1}$ at the nucleus, staying up to 100\,km\,s$^{-1}$ to the NW and decreasing to 75\,km\,s$^{-1}$ to the SE and towards the edges of the FOV. Median radial velocity errors reported by pPXF are 32.6\,km\,s$^{-1}$ in the inner 0\farcs75 and 33.2\,km\,s$^{-1}$ in the inner 1\farcs25. Median errors in the velocity dispersion (also from pPXF) are 30.0\,km\,s$^{-1}$ in the inner 0\farcs75, and 40.2\,km\,s$^{-1}$ in the inner 1\farcs25.

\begin{figure*}[!ht]
\centering
\includegraphics[width=\textwidth, trim={0 13.5cm 0 2cm},clip]{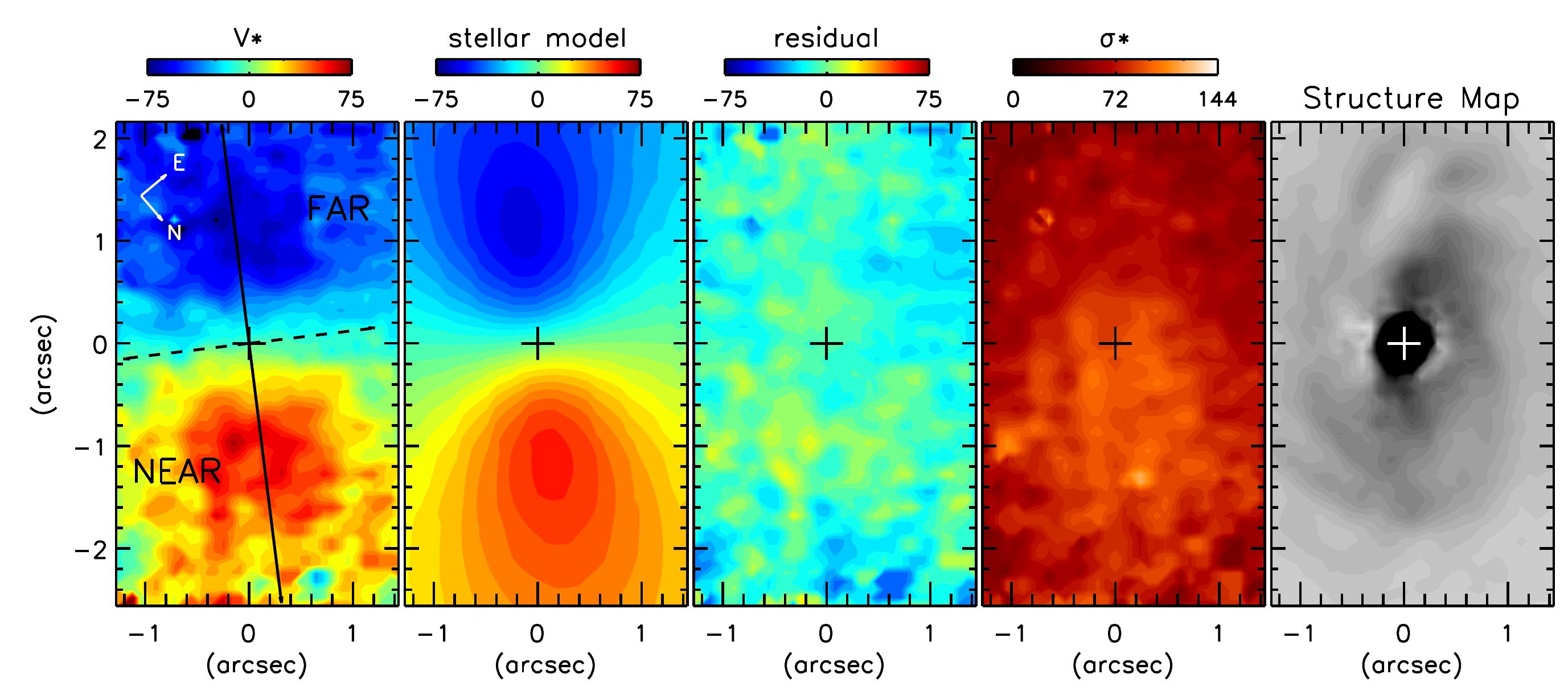}
    \caption{Stellar velocity field, stellar rotation model, residual velocity field (observed - model), stellar velocity dispersion (all in units of km s$^{-1}$ following the colour bars above panels), and the structure map (the bright nucleus masked to show details better at fainter fluxes) for ESO 362-G18. 
In the left panel the solid (dashed) black line indicates the position of the kinematic major (minor) axis. The cross denotes the continuum peak in all panels.}
\end{figure*}
    
    \begin{figure*}[!ht]
\centering
    \includegraphics[width=0.9\textwidth, trim={-1cm 15.7cm 0 0}, clip]{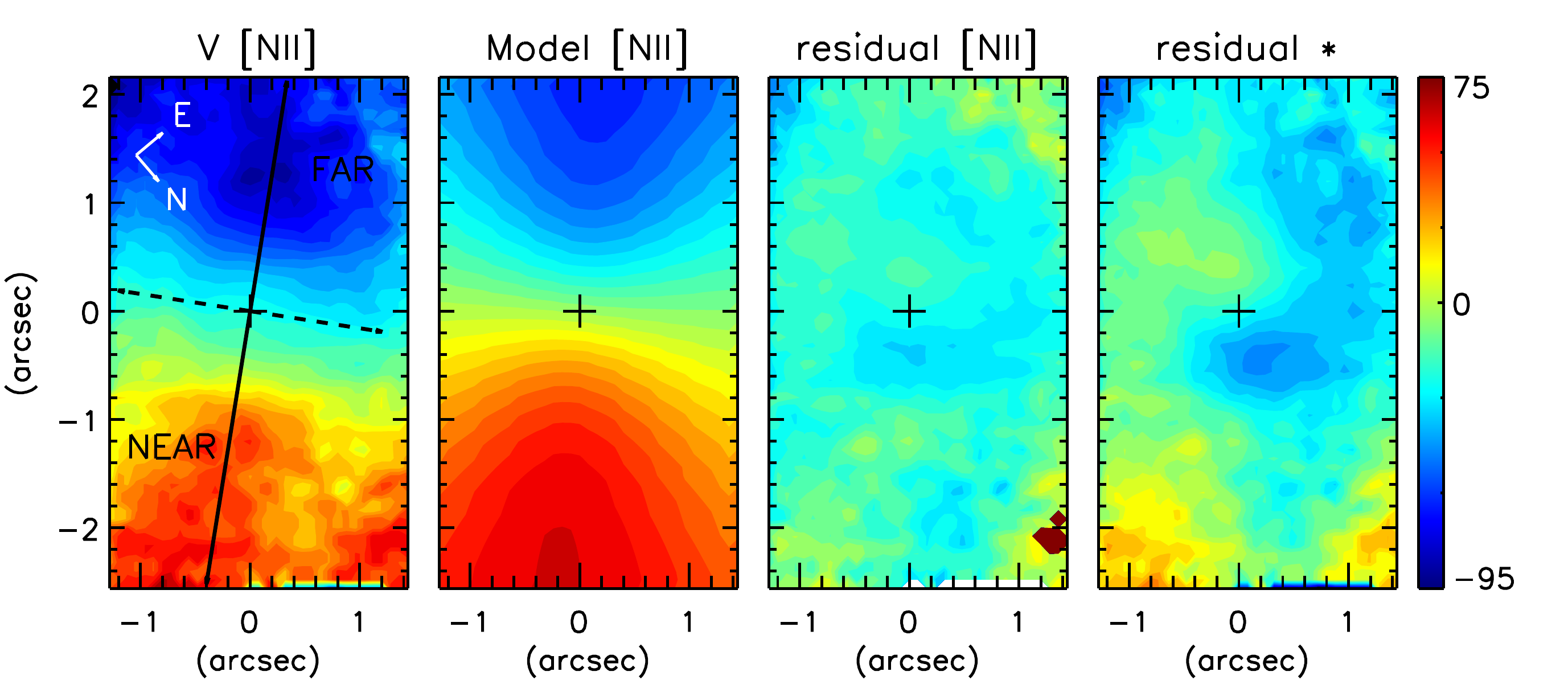}
  \caption{Observed [N\,{\sc ii}] velocity field, [N\,{\sc ii}] rotation model, residual velocity field (observed - [N\,{\sc ii}] model), and residual velocity field (observed - stellar model). The solid (dashed) black line indicates the position of the kinematic major (minor) axis. The cross represents the continuum peak. All panels follow the colour bar (\,km\,s$^{-1}$) shown on the right.}
  \end{figure*}

We employed Kinemetry \citep{Krajnovic2006}, in which the kinematic centre is fixed to the continuum peak, to obtain the PA of the stellar kinematics at various radii from the nucleus. The resulting values range from 130$^{\circ}$ to 139$^{\circ}$, thus we chose the median value of 137$^{\circ}$ as the `global kinematic' PA, as suggested by \citet[Appendix C]{Krajnovic2006}. This value is consistent with the morphological major axes from the literature of between 110$^{\circ}$ and 160$^{\circ}$ \citep[RC3,][and references therein]{Fraquelli2000}.

We model the stellar velocity field by assuming circular orbits in a plane and a spherical potential \citep{Bertola1991}, where the observed radial velocity at a position ($R,\psi$) in the plane of the sky is given by

\small
\begin{equation*}
V=V_{s} + \frac{ARcos(\psi-\psi_{0})}{ [R^{2}[sin^{2}(\psi-\psi_{0})+cos^{2}\theta cos^{2}(\psi-\psi_{0})]+c^{2}cos^{2}\theta ]^{p/2}}
\end{equation*}
\normalsize
where $\theta$ is the inclination of the disk (with $\theta$ = 0 for a face-on disk), $\psi_{0}$ is the PA of the line of nodes measured with respect to the x-axis within the field shown, $V_{s}$ is the systemic velocity, $R$ is the radius and $A$, $c$, and $p$ are parameters of the model. We assumed the kinematical centre to be cospatial with the peak of the continuum emission, a PA of 137$^{\circ}$ as derived via Kinemetry above, a disk inclination of 37$^{\circ}$, obtained from the apparent axial ratio \citep{Winkler1997,Fraquelli2000} under the assumption of a thin disk geometry, and an initial guess of 3731\,km\,s$^{-1}$ for $V_{sys}$ \citep{Paturel2003}. 
We used Levenberg-Marquardt least-squares algorithm to fit the rotation model to the velocity map. The resulting parameters $A$, $c$ and $p$, are 203\,km\,s$^{-1}$, 1\farcs09 and 1.93, respectively. The fitted V$_{sys}$ corrected to the heliocentric reference frame is very similar to our initial guess, so we continue to use the latter. The model stellar velocity field and velocity residuals are shown in Fig. 5.

\subsection{\textbf{Gas kinematics}}

The velocity fields of all strongly detected emission lines, that is 
H$\alpha$, H$\beta$, [N\,{\sc ii}]$\lambda$6583 \text{\AA}, [O\,{\sc i}]$\lambda$6300 \text{\AA}, [O\,{\sc iii}]$\lambda$5007 \text{\AA} and [S\,{\sc ii}]$\lambda$6716 \text{\AA}, show clear signatures of 
rotation; the projected peak rotation velocities range from 66 km s$^{-1}$ up to 82 km s$^{-1}$, although
non-rotation signatures and offset kinematical centres are also present in most lines. The velocity maps from a single Gaussian fit of the [N\,{\sc ii}]$\lambda$6583 \text{\AA}, H$\alpha$, and [O\,{\sc iii}]$\lambda$5007 \text{\AA} emission lines are shown in the left column of Fig. 6 and first and third panels of Fig. 7.

Inspecting the velocity maps of these emission lines, we find an offset of $\approx$ 0\farcs5 (128pc) between the continuum peak and kinematic centre for H$\alpha$, H$\beta$, [O\,{\sc i}], and [O\,{\sc iii}], while no significant offset is present in [N\,{\sc ii}]$\lambda$6583 or [S\,{\sc ii}]$\lambda$6716. To more clearly visualize these offsets, we plot the rotation curves of the stars and the stronger emission lines along their respective kinematic PAs in Fig. 8. The offsets cause an apparent asymmetry in the velocity fields within our FOV, reaching greater blueshifts than redshifts in the majority of cases. However, comparing this feature with previous long-slit spectroscopy of the inner 10\arcsec\ and 30\arcsec\ \citep[respectively]{Bennert2006, Fraquelli2000}, we can infer that this asymmetry is exclusively due to the offset in the kinematic centre.

We used Kinemetry to fit the velocity maps of all emission lines. Given that the rotation curves of 
the emission lines are offset from that of the stars and each other we obtained
reasonable results from Kinemetry only if the kinematic centre was set to a 
position  0\farcs5 to the SE of the continuum peak for H$\alpha$, H$\beta$, [O\,{\sc i}]$\lambda$6300 \text{\AA}, and [O\,{\sc iii}]$\lambda$5007 \text{\AA}; for 
[N\,{\sc ii}]$\lambda$6583 \text{\AA} and [S\,{\sc ii}]$\lambda$6716 \text{\AA},
setting the kinematic centre to the continuum peak gave meaningful results.

Fitting the individual emission line velocity fields with Kinemetry resulted in global kinematic PAs ranging between 121$^{\circ}$ and 139$^{\circ}$. 
For each given emission line, the radial variations of the PA do not exceed 20$^{\circ}$, and all emission line 
global kinematic PAs are in rough agreement (within 16$^{\circ}$) with the stellar kinematic PA except for [S\,{\sc ii}] for which the difference is 24$^{\circ}$. While we expected similar kinematics in H${\alpha}$ and H$\beta$, Kinemetry gives global PAs of 139$^{\circ}$
for H${\alpha}$ and 130$^{\circ}$ for H${\beta}$, and indeed at most radii the fitted PA for H${\alpha}$ is $\sim$9$^{\circ}$ larger than that of H${\beta}$. We thus use a global PA of 134.5$^{\circ}$ for both H${\alpha}$ and H${\beta}$.

\begin{figure*}[!ht]
\centering
    \includegraphics[width=0.9\textwidth, trim={-1cm 15.7cm 0 0}, clip]{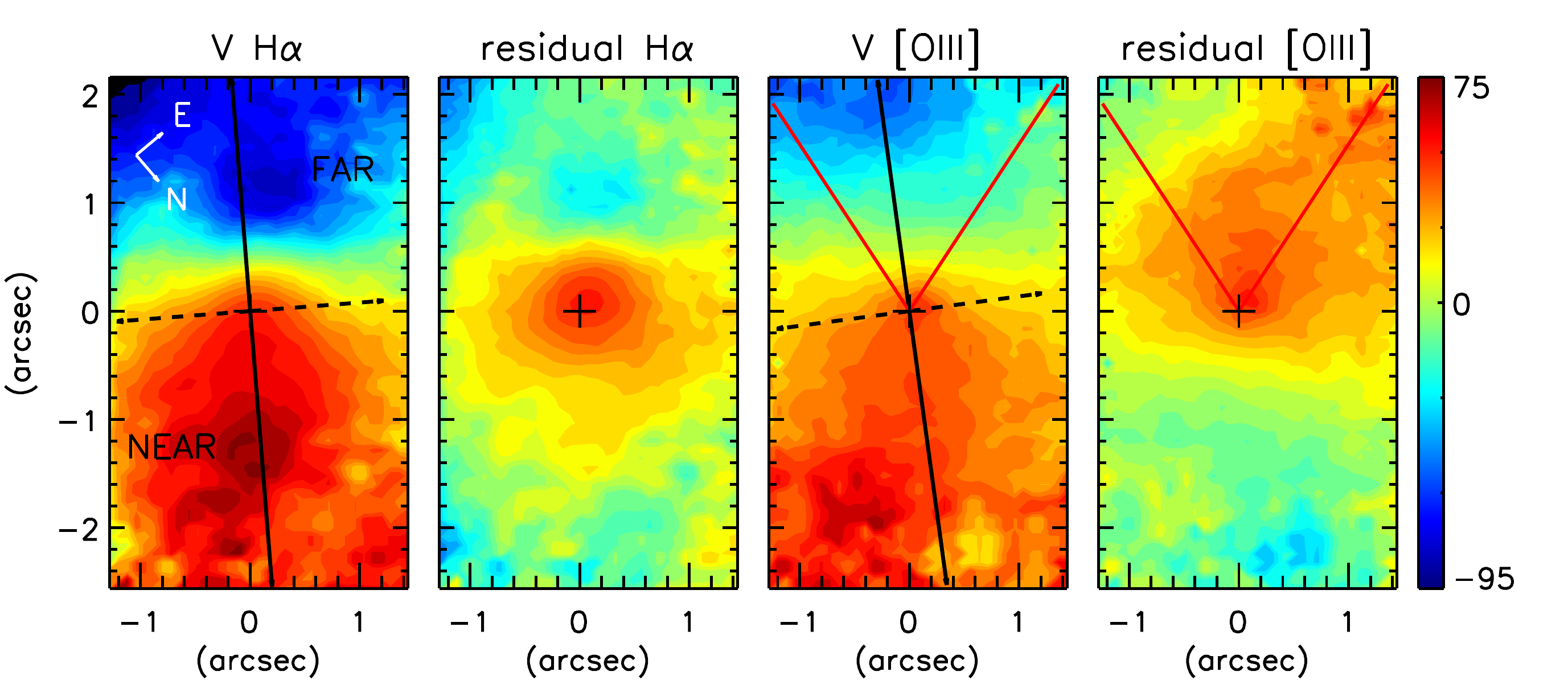}
  \caption{Emission line gas velocity fields and residual velocity fields (observed - [N\,{\sc ii}] model) for H$\alpha$ and [O\,{\sc iii}]. Panels 1 and 3 show the gaseous centroid velocities (km s$^{-1}$) obtained from a single Gaussian fit. The cross denotes the location of the stellar continuum peak, and the solid and dashed lines delineate the major and minor kinematic axes as derived using kinemetry (Section 3.3). For the [O\,{\sc iii}] velocity map (and residual velocity field) the red lines denote the bounds of the ionization cone; we use an opening angle of 70$^{\circ}$ in agreement with the value ($\geq$60$^{\circ}$) proposed by \citet{Fraquelli2000}.
All panels follow the colour bar (km s$^{-1}$) shown on the right.
} 

\end{figure*}

\begin{figure*}[!ht]
 \centering

    \includegraphics[angle=90,width=0.9\textwidth]{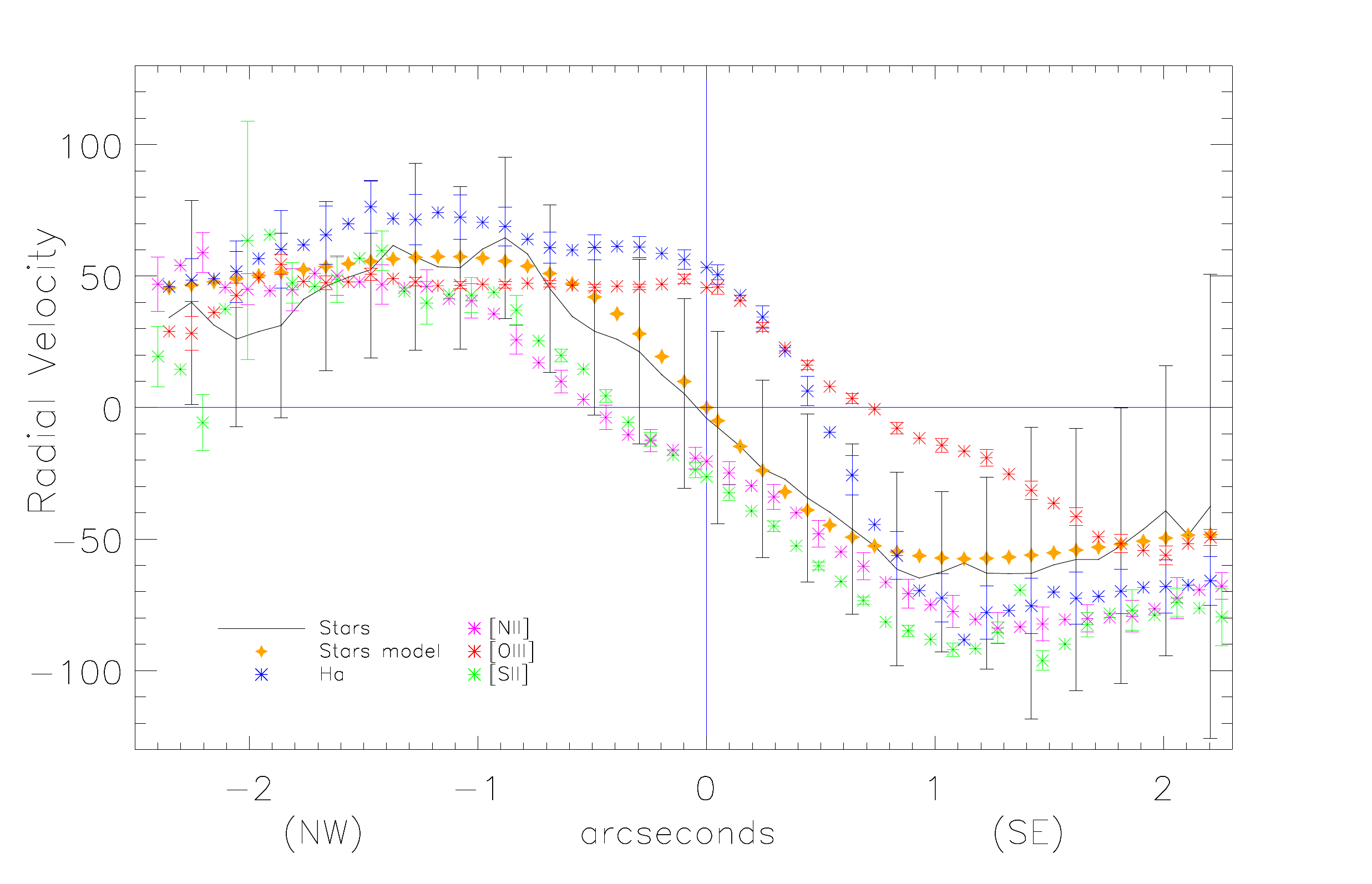}

      \caption{Radial velocity (rotation curves) curves of the emission lines along with the Bertola model (Stars model) for the stellar kinematics, following the legend in the bottom left. 
      The PAs along which these velocities were taken are those derived by Kinemetry (Sect. 3.3) for each emission line: 135$^{\circ}$ for H${\alpha}$, 121$^{\circ}$ for [N\,{\sc ii}], 137$^{\circ}$ for [O\,{\sc iii}] and 113$^{\circ}$ for [S\,{\sc ii}]. The zero velocity (horizontal line) corresponds to a (heliocentric) velocity recession of 3731 km s$^{-1}$. Error bars in the corresponding colour indicate the velocity errors determined by pPXF (stars) or Fluxer (Gaussian fitting to emission lines). The vertical line denotes the position of the continuum peak.}

\end{figure*}




  %




\begin{figure*}[!ht]
\centering
    \includegraphics[width=\textwidth, trim={3cm 18cm -4cm 0},clip]{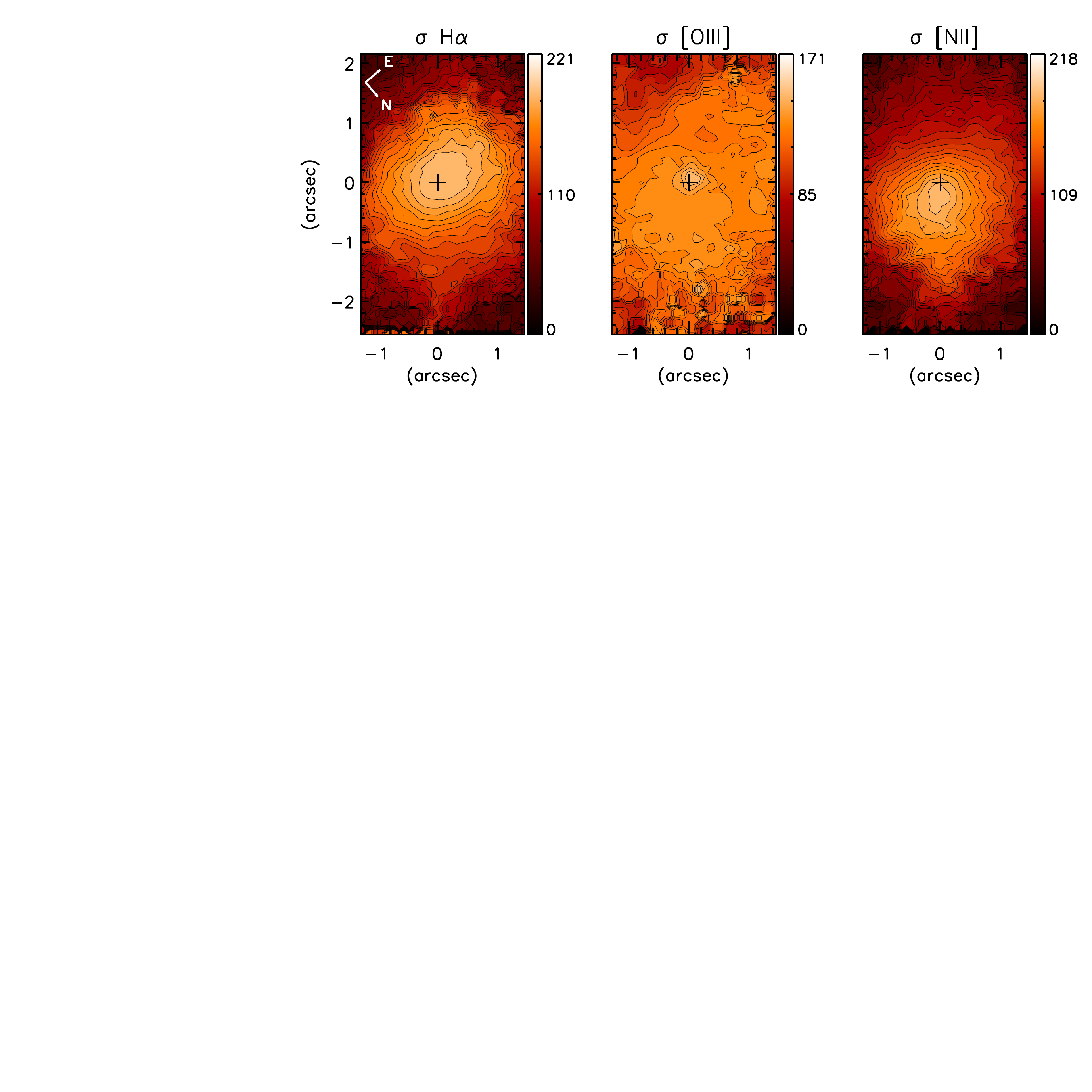}
    \caption{Velocity dispersions (km s$^{-1}$) for H$\alpha$, [O\,{\sc iii}], and [N\,{\sc ii}] emission lines. Each colour bar (km s$^{-1}$) is centred on the average velocity dispersion of the respective line. Crosses denote the continuum peak.}
    
\end{figure*}

Since the [N\,{\sc ii}] emission line is both strong and without a significant kinematic offset from the stellar continuum peak, we fit a gas-kinematic Bertola model to its velocity field, following the procedure outlined in Sect. 3.2. Once more we fix the disk inclination to 37$^{\circ}$, V$_{sys}$  to 3731 km s$^{-1}$, and a kinematic centre cospatial with the stellar continuum peak. The resulting values are 125 km s$^{-1}$, 1\farcs06, 1.04, and 125$^{\circ}$ for A, c, p and the PA, respectively. This gas-kinematics model is shown in the second column of Fig. 6. The third and fourth columns of Fig. 6 show the residual velocity fields of the emission line gas after subtraction of gas-kinematics model and the stellar velocity model, respectively. For all lines except [N\,{\sc ii}] and [S\,{\sc ii}] we see the above-mentioned excess redshift SE of the nucleus. 

Maps of the velocity dispersion (henceforth referred to as $\sigma$ before the name of the corresponding emission line) of the corresponding emission lines, derived with a single Gaussian fit, are shown in Fig. 9. The uncertainty here is $\sim$36 km s$^{-1}$ (instrumental dispersion, $\sigma_{inst}$). The highest nuclear dispersions, $\gtrsim$200 km s$^{-1}$, are seen in $\sigma_{H\alpha}$ and $\sigma_{[N\,{\sc II}]}$; as we leave the nucleus, their dispersions decrease faster along the kinematic major axis than along the kinematic minor axis. On the other hand, $\sigma_{[O\,{\sc III}]}$ is predominantly homogeneous in the non-nuclear regions with its nuclear value rapidly increasing to 170 km s$^{-1}$. In the following sections, we interpret this large nuclear dispersion as due to the presence of a new offset velocity component that is most prominent in the nuclear region causing blended profiles and thus non-reliable fits to a single Gaussian. The dispersion of the H$\beta$ line is similar to that of H$\alpha$, while both $\sigma_{[O\,{\sc I}]}$ and $\sigma_{[S\,{\sc II}]}$ do not present centrally peaked distributions. 

Radial velocity errors were taken directly from FLUXER (Sect. 2). For H$\alpha$, [O\,{\sc iii}] and [N\,{\sc ii}] emission lines, these errors vary between 1 and 24 km s$^{-1}$ in the inner 1\farcs25. As the remaining emission lines observed are not present throughout our FOV, we obtained the errors within the inner 0\farcs75, where they vary in the range between 0.8 and 27 km s$^{-1}$. Errors in the velocity dispersion (also from FLUXER) for H$\alpha$, [O\,{\sc iii}] and [N\,{\sc ii}] emission lines vary between 0.7 and 16 km s$^{-1}$ in the inner 1\farcs25, and between 1.6 and 29 km s$^{-1}$ in the inner 0\farcs75 for the remaining emission lines.

\subsection{\textbf{Position–velocity diagrams}}

To better constrain the emission-line kinematics we built position-velocity (PV) diagrams (Fig. 10) for the three strongest emission lines, [O\,{\sc iii}], H${\alpha}$, and [N\,{\sc ii}]. We centred these PV diagrams on the continuum peak and along PA 130$^{\circ}$ since this is the kinematic major axis found in the single Gaussian fit. The pseudo slit is 0\farcs8 wide. The velocity prediction from the single Gaussian fit is superposed for an easy direct comparison. 

While the PV diagram of [N\,{\sc ii}] shows a good agreement with the single Gaussian fit, the diagrams of both [O\,{\sc iii}] and H${\alpha}$ show a second velocity component redshifted by $\sim$150 km s$^{-1}$ in the nuclear region. 
If the prominent emission in H${\alpha}$ and [O\,{\sc iii}] is seen from a larger velocity range, it may be noted that the most of the emission occurs at velocities below $\pm$ 500 km s$^{-1}$ for both these lines.

\begin{figure*}[!ht]
\begin{subfigure}[b]{0.33\textwidth}
    \includegraphics[width=\textwidth]{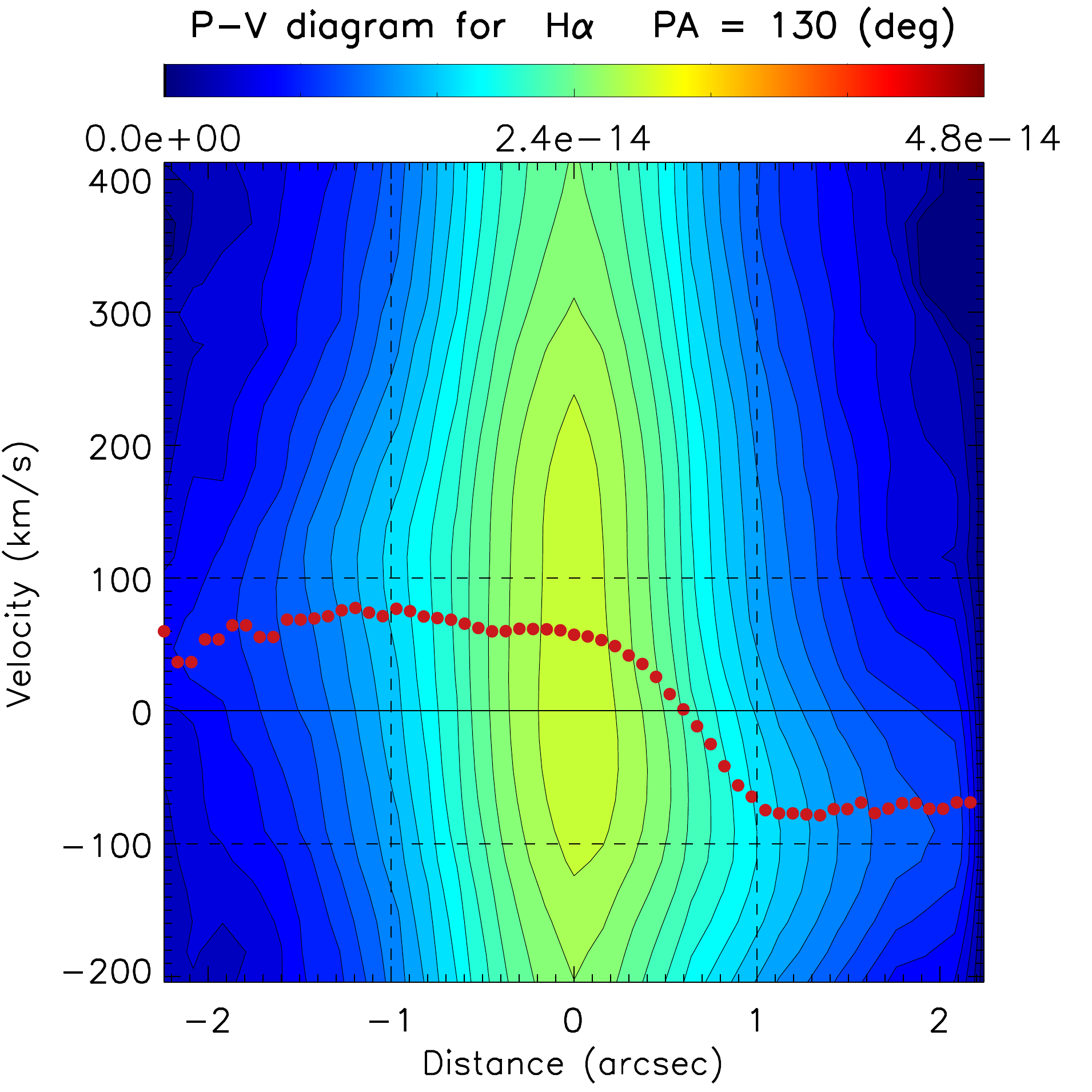}

  \end{subfigure}
  \begin{subfigure}[b]{0.33\textwidth}
    \includegraphics[width=\textwidth]{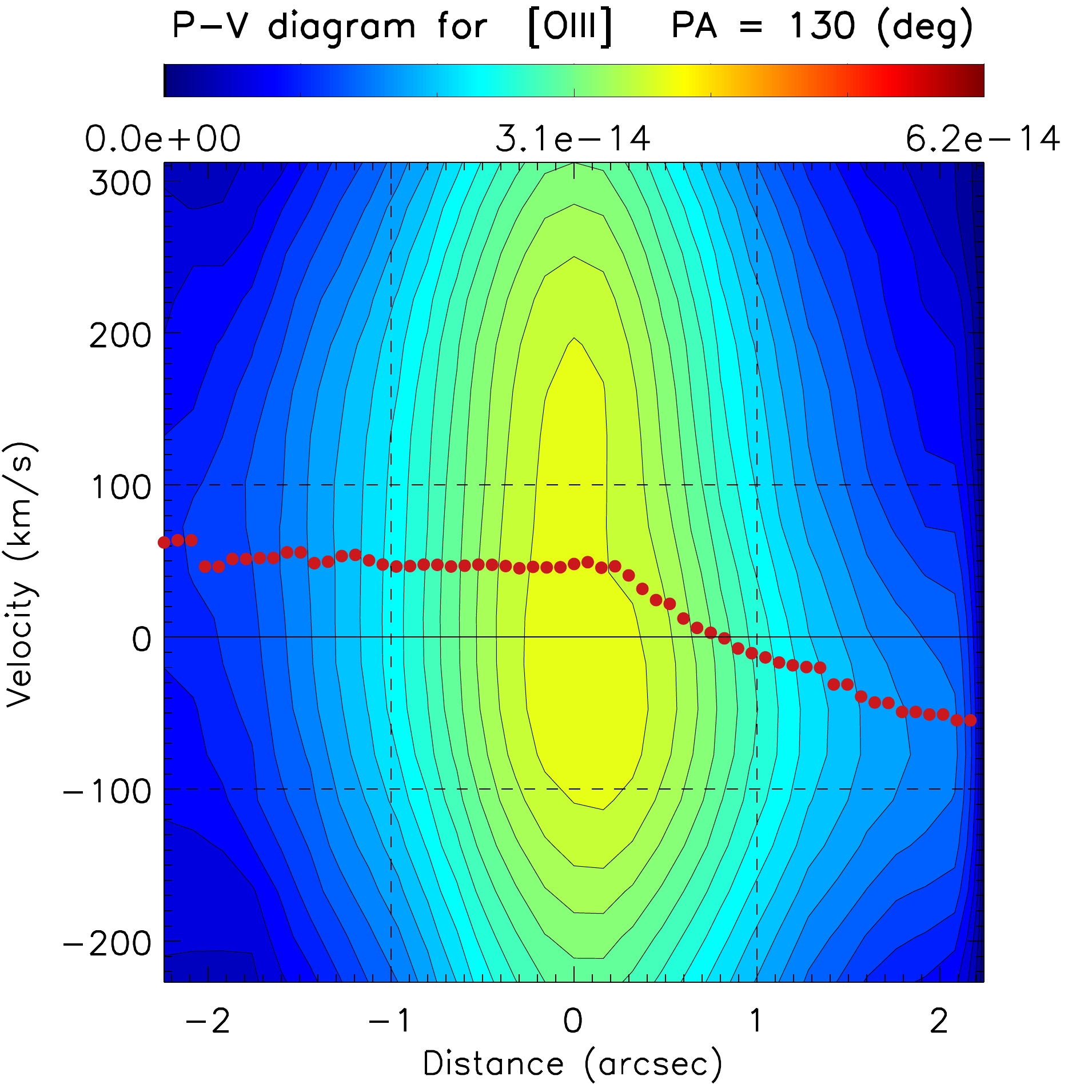}

  \end{subfigure}
  \begin{subfigure}[b]{0.33\textwidth}
    \includegraphics[width=\textwidth]{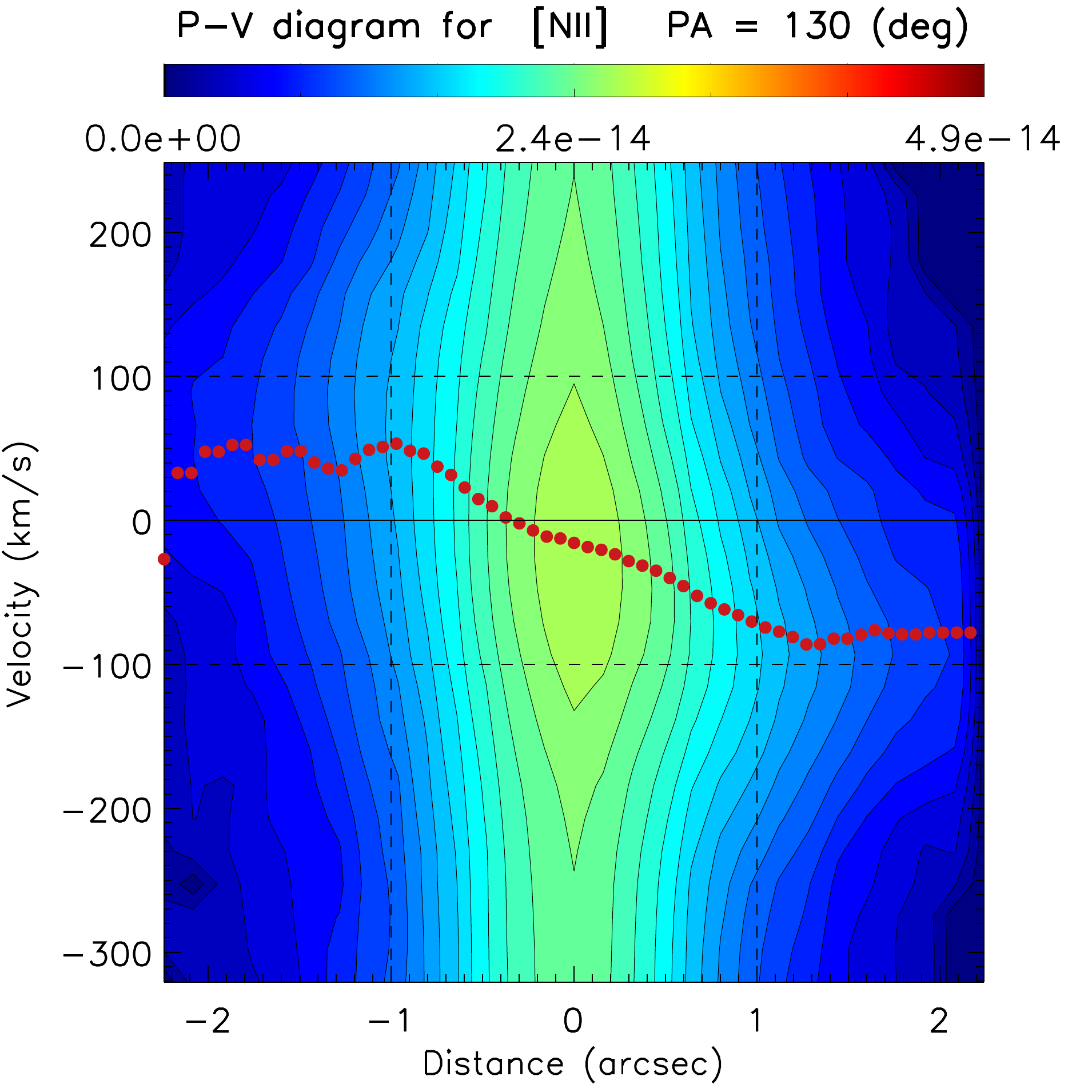}
    
  \end{subfigure}
\caption{From left to right: Position-velocity diagrams of H$\alpha$, [O\,{\sc iii}] and [N\,{\sc ii}] emission lines orientated at PA = 130$^{\circ}$, which is along the kinematic major axis (and the ionization cone). We note that the spectra are continuum subtracted. Colours and contours denote flux (erg cm$^{-2}$ s$^{-1}$ \AA$^{-1}$) following the colour bar above each panel. The corresponding rotation curves obtained from the single Gaussian fit (see Sect. 3.3) are superimposed in red circles. The solid black line indicates the zero velocity (V$_{sys}$ = 3731 km s$^{-1}$), the dashed horizontal lines denotes velocities of $\pm$ 100 km s$^{-1}$, and the dashed vertical lines indicates offsets of $\pm$ 1\arcsec\ from the nucleus. The colour bar above each panel indicates the integrated flux. Contours are in powers of $\sqrt{2}$ to trace both the faint and strong emission structure.
}
\end{figure*}

\begin{figure*}[!ht]
\centering
 \includegraphics[width=0.325\textwidth, trim={1.2cm 0 1cm 0}, clip]{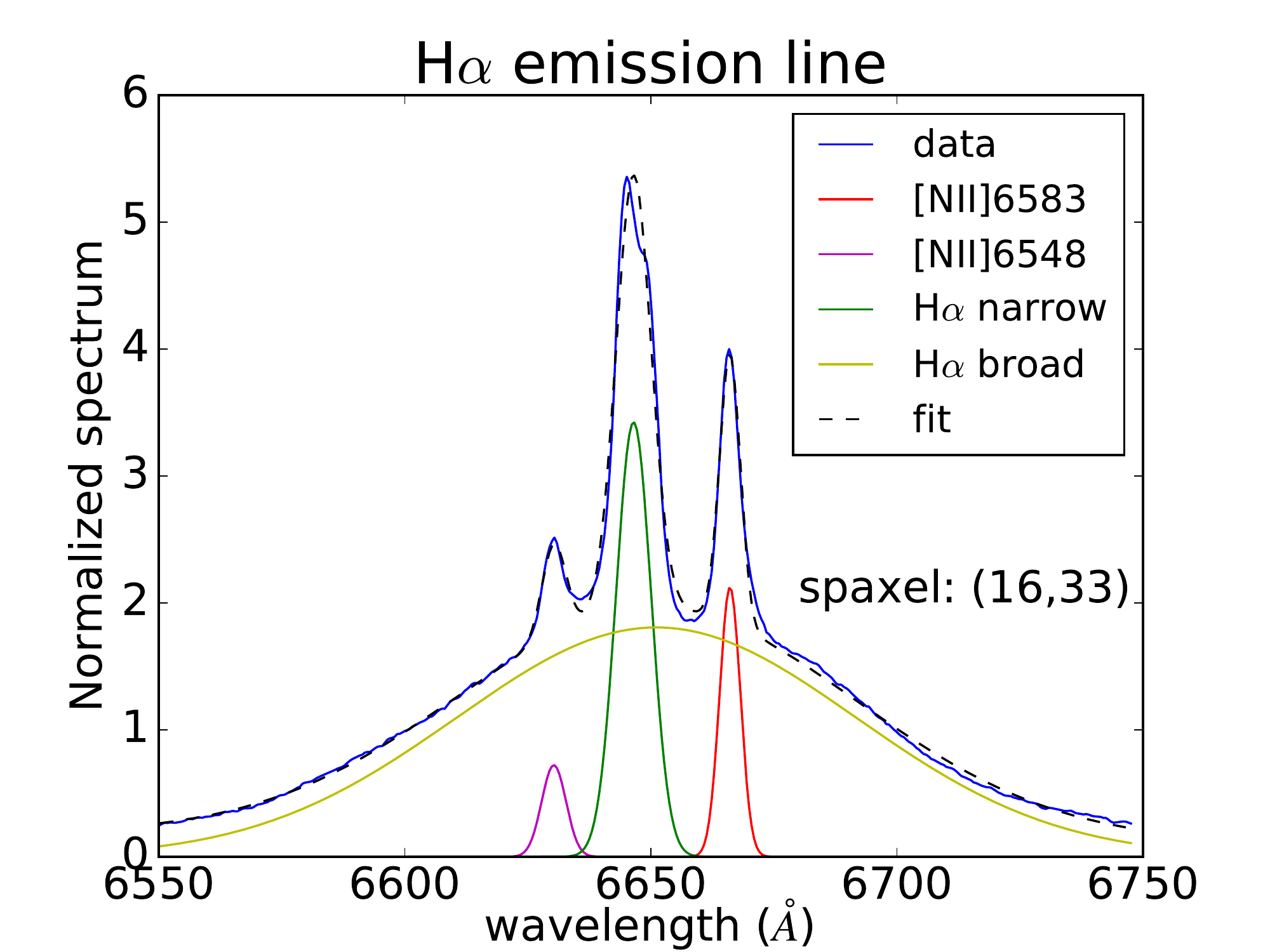}
\includegraphics[width=0.325\textwidth, trim={1.2cm 0 1cm 0}, clip]{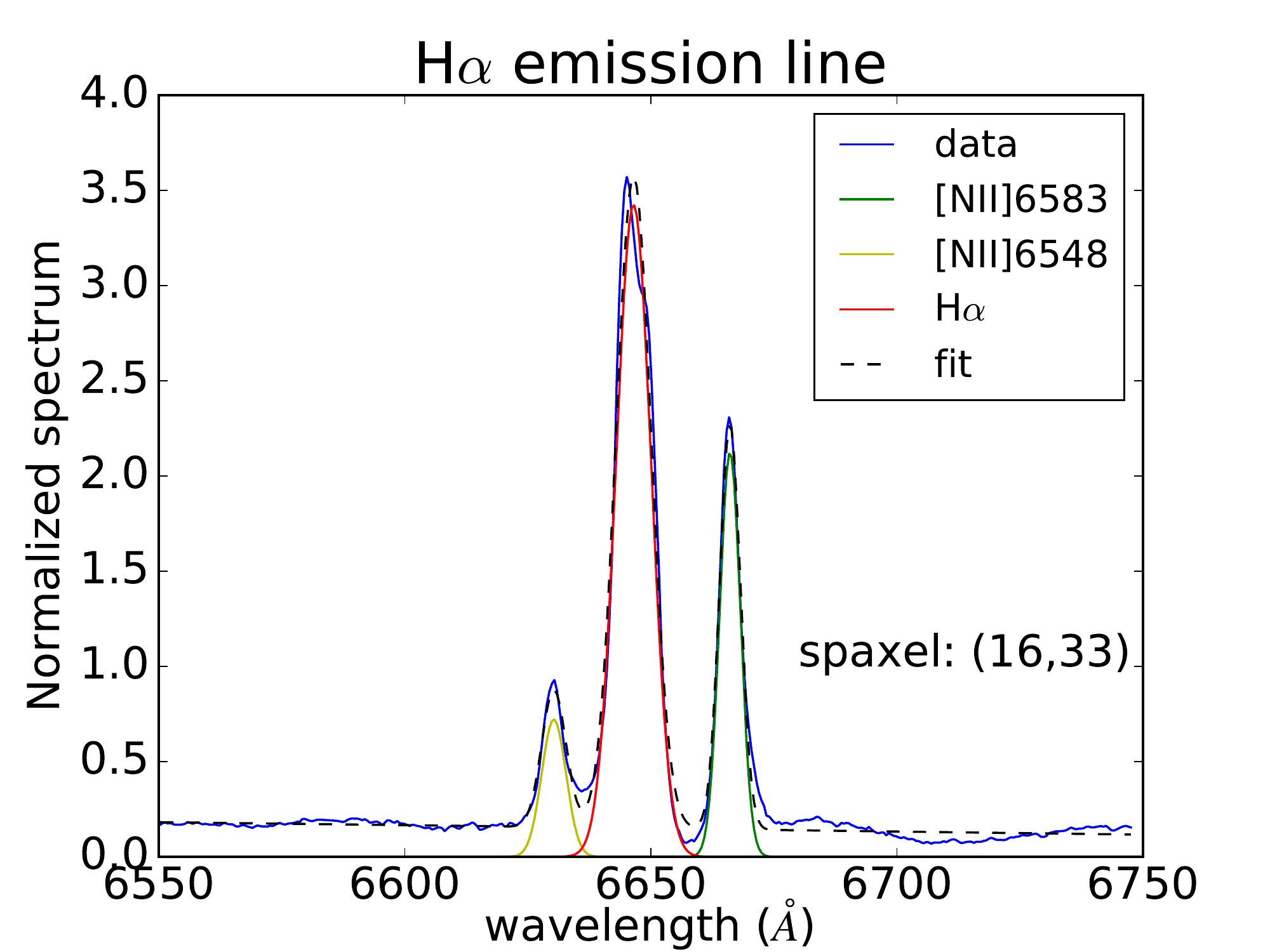}
\includegraphics[width=0.325\textwidth, trim={1.2cm 0 1cm 0}, clip]{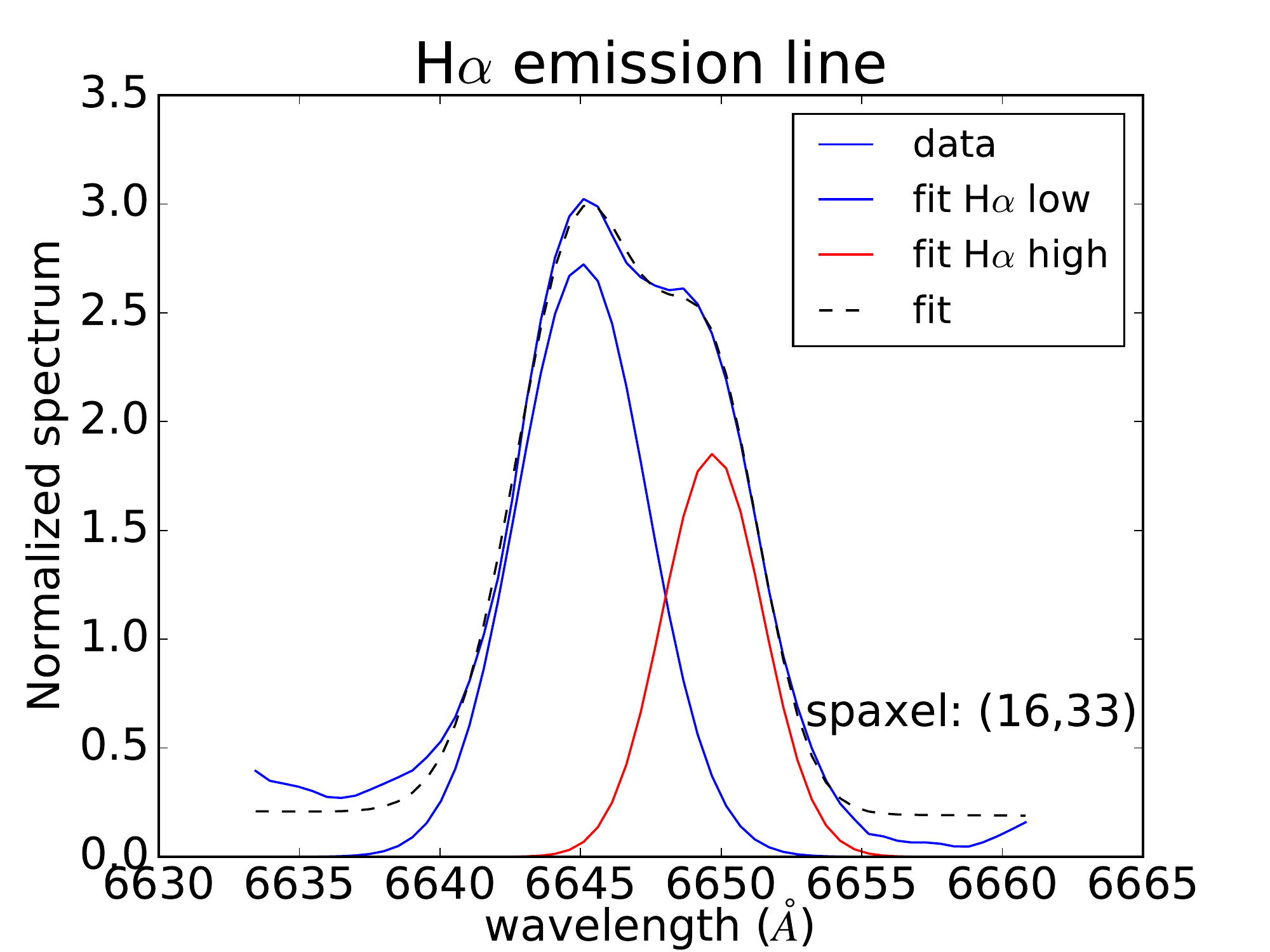}
\includegraphics[width=0.325\textwidth, trim={1.2cm 0 1cm 0}, clip]{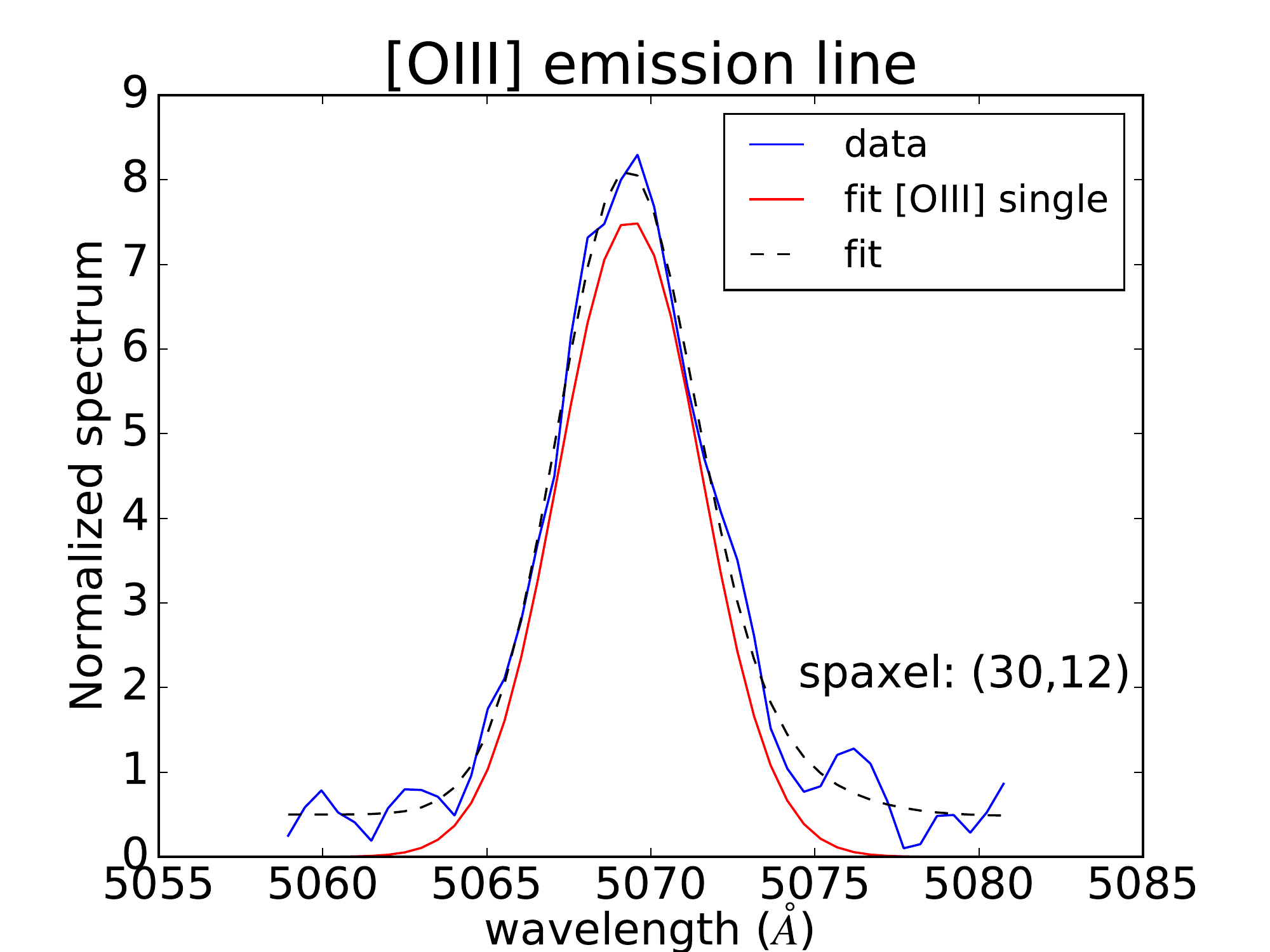}
\includegraphics[width=0.325\textwidth, trim={1.2cm 0 1cm 0}, clip]{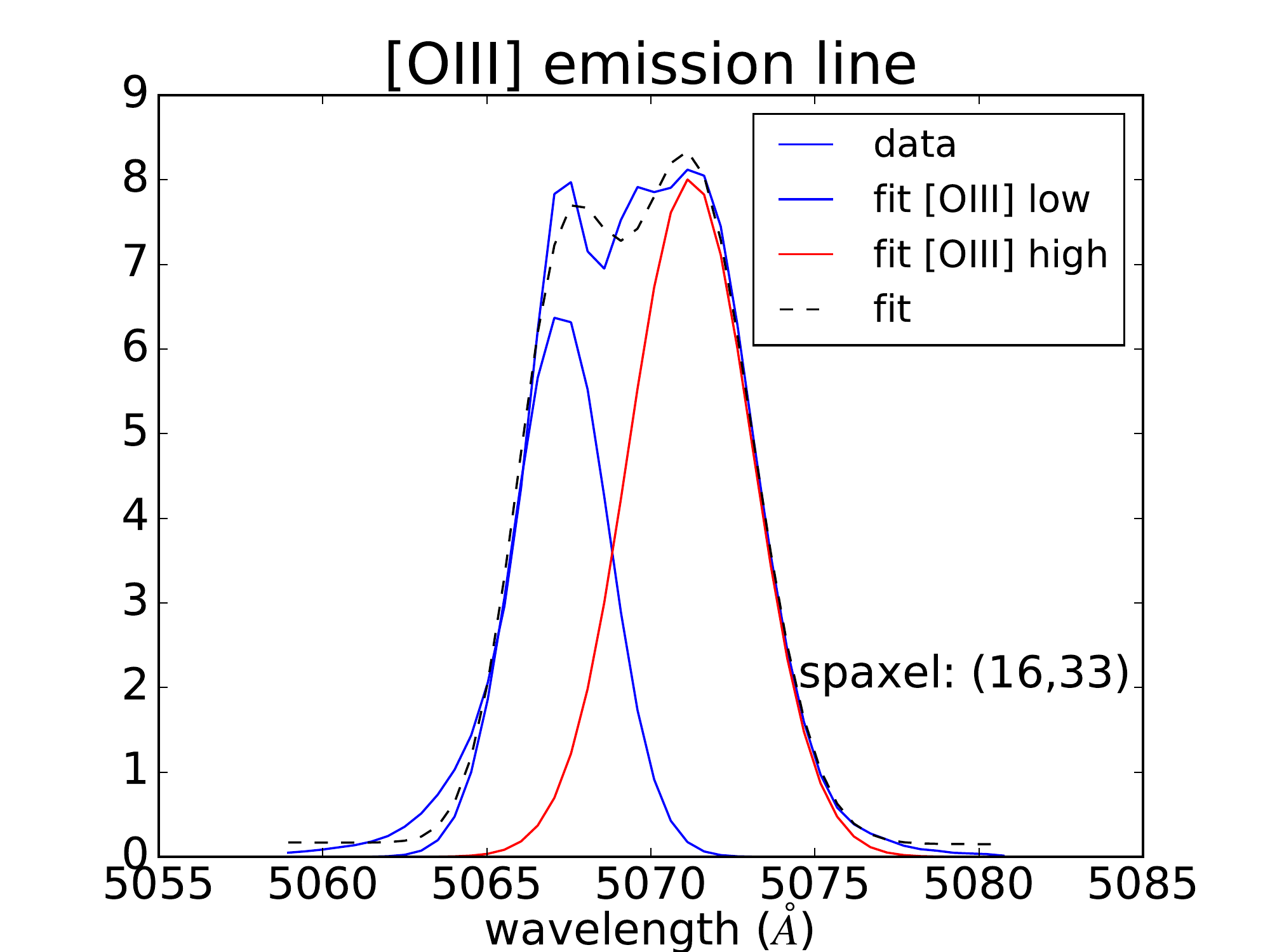}
\includegraphics[width=0.325\textwidth, trim={1.2cm 0 1cm 0}, clip]{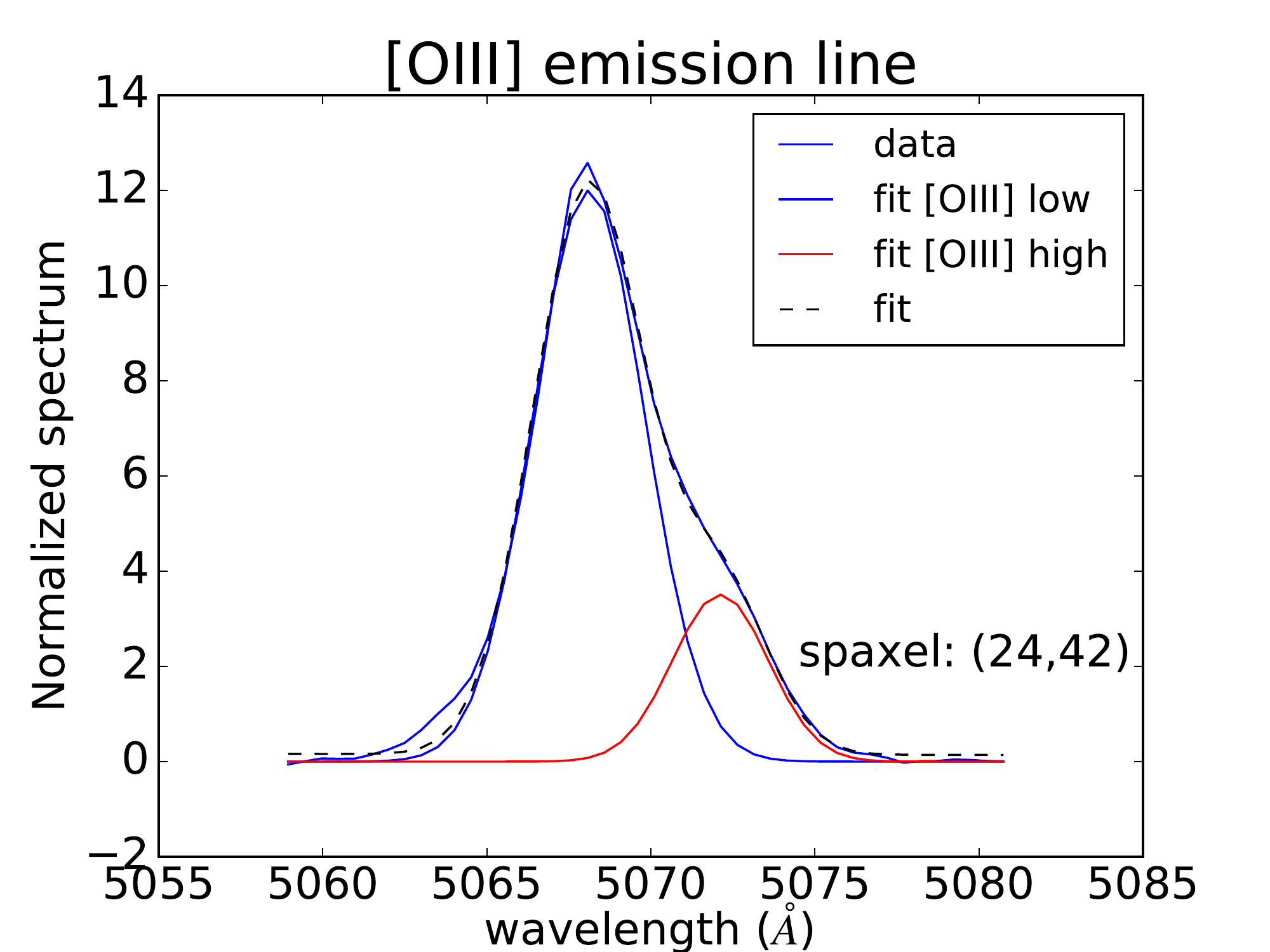}

\caption{Example multiple Gaussian fits to the H$\alpha$+[N\,{\sc ii}] $\lambda \lambda$6548,6583 \text{\AA} (top panels) and [O\,{\sc iii}] $\lambda$5007 \text{\AA} (bottom panels) normalized emission profiles (see Sect. 3.5) for three of the six spaxels shown in Fig. 2. 
In the top panels we show the fitting process used for the H$\alpha$+[N\,{\sc ii}] $\lambda \lambda$6548,6583 \text{\AA} in
a single nuclear spaxel: from left to right, we show: (1) the multiple Gaussian fit to H$\alpha$ (one broad and one narrow
component) and [N\,{\sc ii}] $\lambda \lambda$6548,6583 (one narrow component each) emission lines, (2) the same as
(1) but with the broad H$\alpha$ component subtracted,  and (3) the subsequent double Gaussian fit to the H$\alpha$
line. In the bottom panels, we present examples of Gaussian fits to the [O\,{\sc iii}] emission line in three different
spaxels; the AIC$_{c}$ is used to choose between a single or double Gaussian fit. In spaxel 30,12 (to the N) 
a single Gaussian provides a better fit, while in the spaxels near to the nucleus  (16,33) and  to the E (24,42) double 
Gaussians are required.}
\end{figure*}

\subsection{\textbf{Double Gaussian fit}}

Given the clear evidence for a second velocity component in some of the emission lines, we fit the [O\,{\sc iii}] and H${\alpha}$ emission lines with a double Gaussian; these two lines were chosen as they have the highest S/N ratio among the double peaked lines. Given the clear velocity separation seen between the two components in the PV diagrams, we discriminated these components by their radial velocity (rather than, e.g. width), and they are henceforth referred to as the low-velocity component and high-velocity component. The line profile at each spaxel in the data cube was fit with a double Gaussian using a series of Python codes, mainly within the lmfit package\footnote{https://lmfit.github.io/lmfit-py/intro.html}. To decide whether the observed line profile is a better fit with a single or double Gaussian, we use the corrected Akaike information criterion \citep[AIC$_{c}$,][]{Akaike1974} with the additional caveats that all Gaussian amplitudes are positive. For the H${\alpha}$ emission line, the double Gaussian fit was performed after subtraction of the broad component (Fig. 1), which can be detected as far as $\sim$1\arcsec\ from the nucleus, even though the seeing was $\sim$0\farcs7. This broad component is also present in H${\beta}$ (Fig. 1).

\begin{figure*}[!ht]
 \centering
  \includegraphics[width=\textwidth, trim={0 1cm 0 0}, clip]{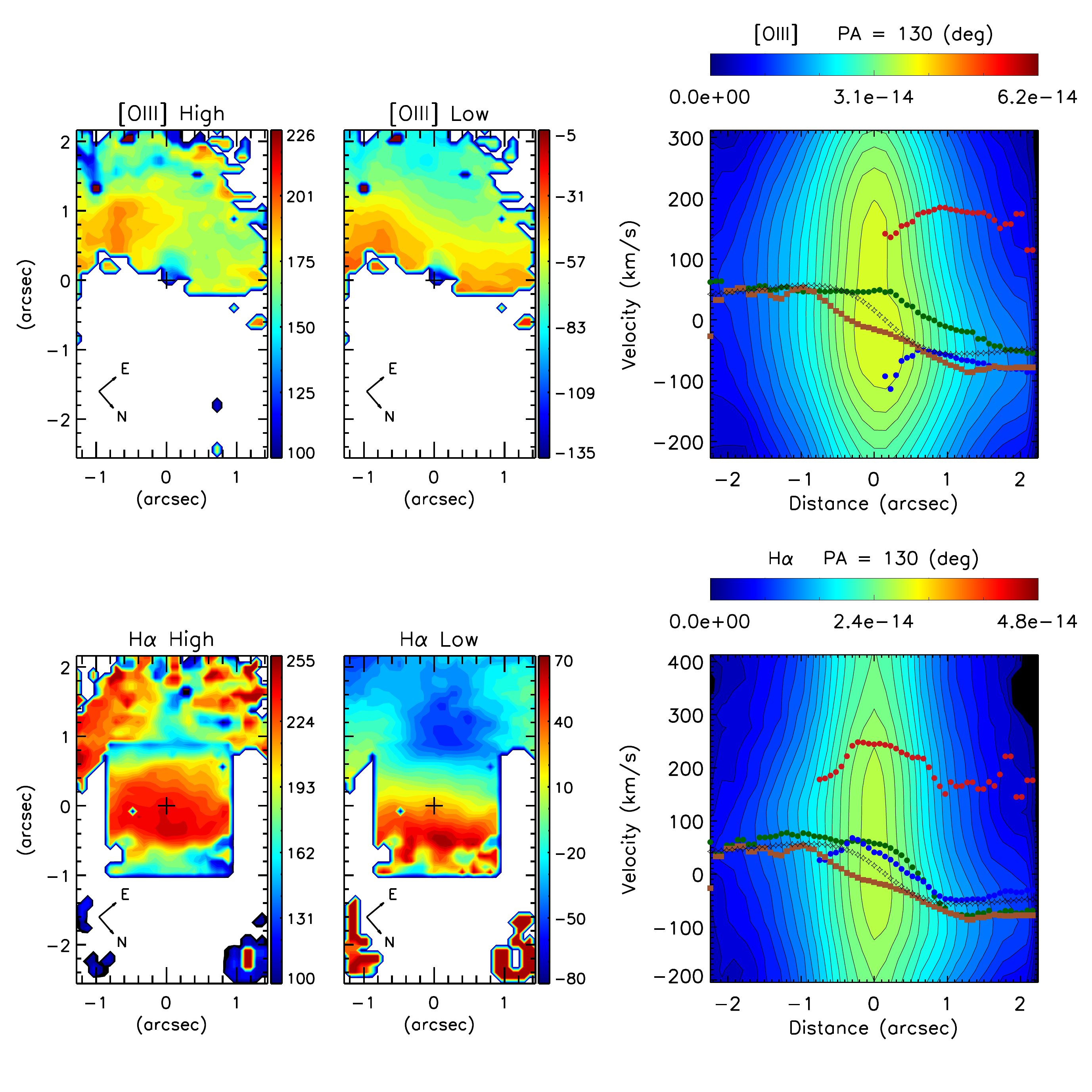}
      \caption{Results of the double component fit to the [O\,{\sc iii}] (top) and H${\alpha}$ (bottom) lines. Left and middle panels: Velocity fields of the high-velocity and low-velocity narrow components, following the colour bar to the right, are shown. White denotes regions where a single Gaussian fit is preferred (see Sect. 3.5). Rightmost panels: PV diagrams as in Fig.~10, but this time we use blue and red points to denote the velocities of the low- and high-velocity components, respectively, and green points to denote the velocities obtained from the single Gaussian fitting. For comparison, we overplot the [N\,{\sc ii}]  (filled brown squares) and stellar (open black stars) velocity curves along the same PA.}
\end{figure*}

The top three panels of Fig. 11 show a detailed example of the fitting process to the H$\alpha$ emission line in a nuclear spaxel. These panels show the multi-component Gaussian fit to the H$\alpha$ and [N\,{\sc ii}] $\lambda \lambda$6548,6583 emission lines, before (left) and after (middle) the subtraction of the broad H$\alpha$ component, and the subsequent double Gaussian fit to the narrow H$\alpha$ emission. In the bottom panels of the same figure, we present examples of single/double (as decided by the AIC$_{c}$) Gaussian fits for [O\,{\sc iii}]  in different spaxels (($x$,$y$) axes) of our data cube. 

To estimate the errors in the velocities of the narrow components of H$\alpha$ produced by an erroneous broad component subtraction, we use the following iterative process. For each spaxel, we vary the central velocity of the broad line over the range $\pm3$ times the $1\sigma$  velocity error reported by $lmfit$. The double narrow components are then fit after subtraction of this broad line, and the results compared to those of the best fit 
(Fig. 12, bottom panels). We find that the narrow component velocities vary by less than 1.8 km s$^{-1}$ for the low-velocity  component and typically less than 6 km s$^{-1}$ for the high-velocity component (in this latter two spaxels show velocity differences of up to 38 km s$^{-1}$). The errors recorded by lmfit for the narrow H$\alpha$ component velocities are negligible. Less than 3\% of the spaxels which were originally better fitted with a double (instead of single) Gaussian are occasionally better fitted with a single Gaussian during our iterative process. Thus, overall, the two narrow component velocities are robust w.r.t. the seeing-smeared contamination of the (unresolved) BLR. 

Does the velocity profile of the unresolved (but seeing-smeared) nuclear NLR also affect the velocities of the two-component Gaussian fit? It is difficult to quantify this effect, but we note that the PV diagrams of [O\,{\sc iii}] and H${\alpha}$ (see Figs. 10 and 12 for those along the major axis) 
clearly show that the velocity profiles are not symmetric about the nucleus even in the nuclear seeing disk. The velocities from the double component Gaussian fit appear to relatively well trace features seen in the PV diagrams, so any unresolved nuclear component would smooth out but not hide variations. We note however that to the NW of the nucleus, the two-component fit to the H${\alpha}$ line does not appear to follow the PV diagram. Instead the PV diagram is more consistent with a single Gaussian at a slightly larger redshift.

\begin{figure*}[!ht]
 \centering
    \includegraphics[width=\textwidth, trim={1cm 18cm -2cm 0},clip]{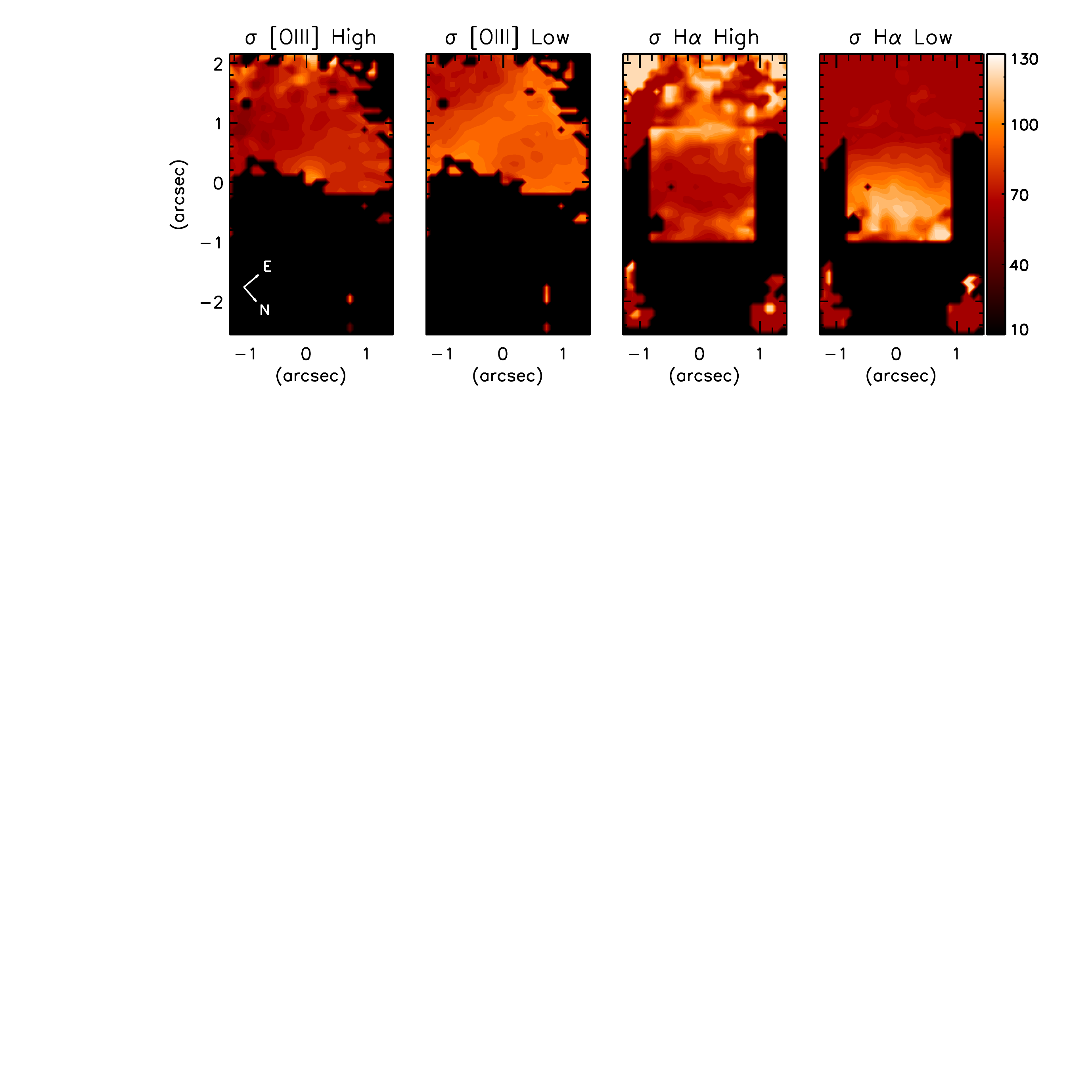}
      \caption{From left to right: The velocity dispersion of the [O\,{\sc iii}] and H${\alpha}$ high-velocity and low-velocity components, respectively. All panels follow the colour bar (km s$^{-1}$) shown on the right.}
\end{figure*}

The results of the two-component fits to [O\,{\sc iii}] and H${\alpha}$ are shown in  Fig. 12. The left four panels show the velocity fields of the two components of H$\alpha$ and [O\,{\sc iii}], and the two rightmost panels of Fig. 12 show the same major axis PV diagrams as Fig. 10 (along PA 130$^{\circ}$). But this time we overlay the velocities of each of the two velocity components (high-velocity component in red and low-velocity component in blue) along with the velocities obtained from the single Gaussian fit to the respective line (green), the velocities obtained by a single Gaussian fit to the [N\,{\sc ii}] line (brown) and the Bertola model fit to the stellar kinematics (black). According to the AIC$_{c}$, the two-component fit is required only to the SE. To the NW, a single component fit gives better results. The lower [O\,{\sc iii}] velocity component has velocities ranging from -30 km s$^{-1}$ to -135 km s$^{-1}$ and a kinematic PA of $\sim$ 123$^{\circ}$, while the higher [O\,{\sc iii}] velocity component shows values $\approx$ 200 km s$^{-1}$ higher. The lower H${\alpha}$ velocity component shows velocities of -80 km s$^{-1}$ to 70 km s$^{-1}$, with a PA of $\sim$ 123$^{\circ}$, while the higher H${\alpha}$ velocity component show values of 100 km s$^{-1}$ up to 255 km s$^{-1}$. 

The corresponding velocity dispersions are shown in Fig. 13. The low-velocity component of H${\alpha}$ shows a centrally peaked velocity dispersion map, while in the higher velocity component shows high dispersions only in disjoint regions $\gtrsim$0\farcs6 from the nucleus. The low-velocity component of [O\,{\sc iii}] has systematically higher dispersions (except in the nucleus) in comparison to the high-velocity component.

\subsection{\textbf{Black hole mass}}

Black hole mass estimations for ESO 362-G18 were rigorously explored by \citet{AgisGonzalez2014}; our new observations allow us a better constraint on the value of the FWHM of $H{\beta}$ since we have a better spectral resolution and two-dimensional data for this line. We fit the $H{\beta}$ profile with a double Gaussian using the lmfit package as described in Sec. 3.5, and use the FWHM  of the broad component (hereafter FWHM$_{H{\beta}}$) to estimate the black hole mass.

Assuming a disk-like BLR geometry to avoid assuming a virial coefficient $f$, which can vary widely, \citet{AgisGonzalez2014} used the following expression:

\begin{equation*}
M_{BH} = R_{BLR} FWHM_{H{\beta}}^{2}  (4G sin^{2} i )^{-1}
\end{equation*}

\noindent
where $i$ is the angle between the LOS and the angular momentum vector of the disk-like BLR (53$^{\circ}$ $\pm$ 5$^{\circ}$), R$_{BLR}$ is the radius of the BLR ($\sim$ 5.2 $\times$ 10$^{16}$ cm), and $G$ is the gravitational constant. Using our new value of FWHM$_{H{\beta}}$, 5689$^{+398}_{-723}$ km s$^{-1}$, calculated in the inner $\sim$0\farcs35 around the continuum peak (i.e. within the seeing disk), we obtain a black hole mass $M_{BH}$ of 4.97$^{+1.60}_{-1.61}$ $\times$ 10$^{7}$ M$_{\odot}$, which is consistent with the value (4.5 $\pm$ 1.5 $\times$ 10$^{7}$ M$_{\odot}$) obtained by \citet{AgisGonzalez2014}, and in a range typical of both narrow line and broad line Seyfert 1 galaxies \citep[][Fig. 4]{Greene2005}.

\section{Discussion}

The structure map (rightmost panel of Fig. 5) shows spiral arms that get increasingly fainter as they approach the nucleus. Given the scale of the FOV and the asymmetries we find after subtracting two-dimensional Gaussians from the flux maps of each emission line, we interpret the unusual spiral arm structure as a result of instabilities produced by the inner Lindblad resonance, which is expected to lie in the inner 1.6 kpc \citep{Laine2002}. We thus argue that the unusually high value of the parameter p ($>$1.5) in the Bertola stellar model is a direct consequence of this resonance. The abundant dust seen in the structure map also supports the hypothesis of \citep{Simoes2007} that the presence of dust is a necessary condition for accretion onto the nuclear SMBH.

Previous studies of ESO~362-G18 have found an asymmetric [O\,{\sc iii}] emission-line morphology with a fan-shaped structure extending 10\arcsec\ from the nucleus to the SE, and an asymmetric morphology in the stellar continuum that is reminiscent of a minor merger system \citep[Fig. 29]{Mulchaey1996}. \citet{Fraquelli2000}, using long-slit spectra, noted that the kinematics of this extended emission-line region is similar to that of the stars. They posited, primarily based on the morphological appearance, the presence of an AGN outflow with a collimating axis orientated at an angle $\leq$30$^{\circ}$ with respect to the galactic plane; this small angle is required to allow the nuclear radiation to intercept the gas in the disk and to allow a direct LOS to the BLR. They proposed an opening angle larger than 60$^{\circ}$ for the ionizing radiation cone. We note that the spectral resolution of the long-slit spectra used by \citet{Fraquelli2000} was not high enough to resolve the nuclear outflow that we posit. Our results are consistent with the outflow scenario posited by \citet{Fraquelli2000}: both [N\,{\sc ii}] and [S\,{\sc ii}] share the same kinematic PA ($\approx$120$^{\circ}$), $\approx$10$^{\circ}$ less than the kinematic PA of H$\alpha$, H$\beta$, [O\,{\sc i}], and [O\,{\sc iii}] ($\approx$130$^{\circ}$).
The offset in the kinematic centres of H$\alpha$, H$\beta$, [O\,{\sc i}] and [O\,{\sc iii}] are closest to the direction of the ionization cone \citep[158$^{\circ}$,][]{Fraquelli2000}. Both facts allow us to infer that the high-ionization emission lines are more affected by this cone, while [N\,{\sc ii}] and [S\,{\sc ii}] appear to be dominantly from gas rotating in the galactic plane, and following a rotation curve similar to that of the stars (Fig. 8).  

The PV diagrams (Fig. 10) of [O\,{\sc iii}] and H${\alpha}$ clearly show a second velocity component $\sim$150 km s$^{-1}$ to the red. Its contribution is most significant in the inner arcsecond and is the reason why the single Gaussian fit gives velocities redder than the expectation of pure rotation in the nucleus (see Fig. 7). 
The equivalent PV diagrams for [O\,{\sc i}] and H${\beta}$ (not shown) are also consistent with the presence of the second higher velocity component, but we could not rigorously fit double Gaussians to these profiles owing to their relative faintness, especially at distances $\gtrsim$ 1.5\arcsec\ from the nucleus (see Fig. 1). Further, both [O\,{\sc i}] and H${\beta}$ (not shown) show the same asymmetries as [O\,{\sc iii}] and H${\alpha}$ in their velocity maps derived from a single Gaussian fit (Fig. 7).

The appearance of the velocity fields (PAs, velocity ranges, and rotation curves) of the low-velocity component of [O\,{\sc iii}] and H${\alpha}$ (middle panels of Fig. 12) are very similar to those derived from [N\,{\sc ii}] in the same region. Therefore we interpret this component as emission from gas in the galactic disk that is rotating in the same manner as the [N\,{\sc ii}]-emitting gas and the stars. Only the negative velocities in [O\,{\sc iii}], reached very close to the nucleus (from -75 km s$^{-1}$ to -120 km s$^{-1}$), can be attributed to an outflow approaching the observer.

The high-velocity component of both these lines shows a very different velocity field, with values that exceed 200 km s$^{-1}$ ($>$ 330 km s$^{-1}$ deprojected). We thus conclude that the high-velocity component corresponds to the bright gas within the AGN ionization cone. Given that (deprojected) velocities larger than 150 km s$^{-1}$ are typically observed in outflows instead of inflows in nearby galaxies (see \citet{Barbosa2009}, \citet{Storchi-Bergmann2010} and \citet{Riffel2011} for examples of outflow velocities, and \citet{Fathi2006}, \citet{SchnorrMuller2011} and \citet{SchnorrMuller2014b} for examples of inflow velocities), the most plausible explanation is that the high-velocity component is gas entrained by the AGN outflow at an angle \textit{i} greater than zero, located behind the plane of the sky from our LOS, and thus redshifted to the observer. Why would we preferentially see gas on the far side of the ionization cone (redshifted to us) rather than on the near side (which would be blueshifted)? The explanation lies in the illumination of the NLR clouds by the AGN: on the far side of the cone in ESO 362-G18 we see the side of the gas cloud that is illuminated by the AGN, while on the near side we see primarily the dark side of the NLR clouds \citep[see e.g. ][]{Lena2015}. The low- and high-velocity components in ESO 362-G18 are reminiscent of the case in NGC 4151 \citep{Storchi-Bergmann2010}. The difference is that in NGC 4151 the high-velocity component corresponds to gas illuminated by a symmetric bicone, extending both in front and behind the galactic plane, while in ESO 362-G18 only the gas in front the galactic disk, illuminated by a single ionization cone, is seen.

  \citet{AgisGonzalez2014} have estimated an inner accretion disk inclination of 53$^{\circ}$ $\pm$ 5$^{\circ}$; on the other hand, using the ratio of the minor to major photometric axis from \citet{Winkler1997}, \citet{Fraquelli2000} derived a galactic disk inclination of $\approx$ 37$^{\circ}$. Therefore, we suggest a picture for ESO 362-G18 in which the ionization cone has an inclination angle \textit{i} $\sim$ 8$^{\circ}$ $\pm$ 5$^{\circ}$ with respect to the plane of the sky with a half-opening angle of 45$^{\circ}$ in such a way that the cone intersects with the galactic disk in the SE direction, illuminating gas receding from our LOS due to the outflow; this value of the half-opening angle of the ionization cone was also suggested by \citet{AgisGonzalez2014}. We only see blueshifted gas very close to the nucleus (see Fig. 12, middle top panel), where the [O\,{\sc iii}] gas is entrained by the approaching side of the cone in a small region corresponding to the thickness of the disk or bulge of the galaxy. This proposed configuration for the nuclear region in ESO 362-G18 is shown schematically in Fig. 14.
  
 For both [O\,{\sc iii}] and H${\alpha}$, the highest dispersion values are predominantly reached in the low-velocity component (Fig. 13), which we interpret as gas rotation in the galactic disk. Only the central region of the high-velocity component (the outflow component) in [O\,{\sc iii}] shows dispersions higher than 105 km s$^{-1}$, which we can interpret as coming from the approaching side of the outflow where the outflow is still within the galactic disk. The value of $\sigma$$_{H{\alpha}}$ of the outflow component is very sensitive to the subtraction of the broad emission in H$\alpha$. This process was carried out before the two Gaussian fit, so the large dispersion values ($\sim$ 0\farcs8 SE from the nucleus) should be taken with some reserve since this is potentially due to confusion with the broad emission. We can only deduce that the decrease of $\sigma$$_{[O\,{\sc iii}]}$ and $\sigma$$_{H{\alpha}}$ in the outflow component is due to its partial occultation by the galactic disk.

\begin{figure}[!ht]
 \centering
 \includegraphics[trim={0 1cm 0 0},width=0.5\textwidth]{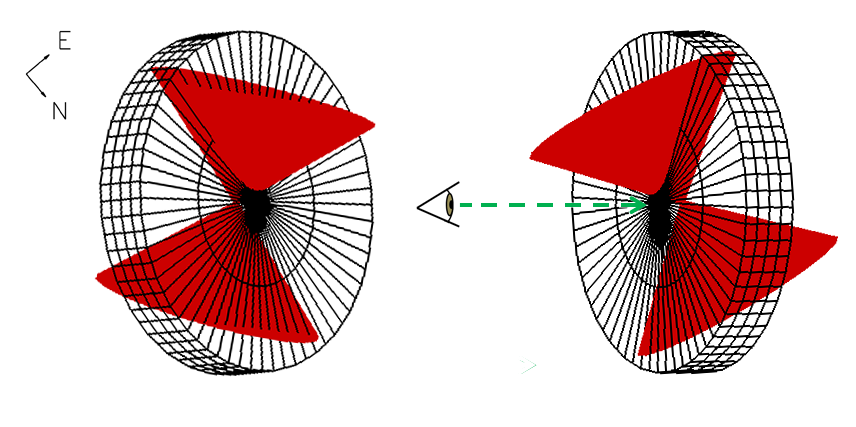}
       \caption{Proposed configuration for the nuclear region of ESO 362-G18 (see Section 4) where the galactic disk is shown both from our perspective (left; in the same orientation as the data cube) and in profile (right). The SE (top) ionization cone is towards us and intersects the galaxy disk primarily behind the plane of the sky, thus generating the high-velocity (redshifted) component of [O\,{\sc iii}] and H${\alpha}$. This cone also intersects the galaxy disk in front of the plane of the sky, but only in a very small region close to the nucleus, producing the greatest blueshift seen in the low-velocity component of [O\,{\sc iii}]. The NW (bottom) ionization cone, if it does exist, is almost always hidden by the galactic disk for our LOS.}
\end{figure}

Since the outflow velocities posited above are not as high as the dividing line commonly adopted to discern immediately between starburst-driven superwinds and AGN-driven outflows \citep[$>$500 km s$^{-1}$,][]{Fabian2012,Cicone2014}, we must test whether the kinetic power injected by supernovae in the inner 100~pc is sufficient to drive the outflow. The outflow, while it extends over at least $\sim$0.5 kpc to the SE, is already seen 
at high velocities in the inner seeing disk ($\sim$0.1~kpc). The global SFR of ESO 362-G18 is relatively low: 
the H$\alpha$-derived global SFR is $\sim$ 5.5 $\times$ 10$^{-3}$ M$_{\odot}$ yr$^{-1}$ \citep{Galbany2016} and we 
derive a similar value from the far-infrared luminosity (based on IRAS fluxes). We note that  \citet{Melendez2008} 
quoted a relatively high instantaneous SFR of 0.85 M$_{\odot}$ yr$^{-1}$ in the inner kpc, based on Spitzer spectroscopy of the 
[Ne\,{\sc ii}] line; these authors, however, made many estimations and assumptions beyond those previously used \citep[e.g.][]{Genzel1998,Ho2007} in the disentanglement of the AGN versus star formation contribution to the [Ne\,{\sc ii}] line and its posterior conversion to a SFR method. Given that the outflow is seen in the inner kpc and starts within the inner $\sim$100~pc, the SFR in this corresponding nuclear region is expected to be significantly lower than 10$^{-2}$ M$_{\odot}$ yr$^{-1}$.

Using the relationship of \citet{Veilleux2005}, $P_{kin,SF}$(erg s$^{-1}$) = 7$\times$ 10$^{41}$ SFR(M$_{\odot}$ yr$^{-1}$), the galaxy-wide kinetic power injected by supernovae, $P_{kin,SF}$, is  $\sim$ 10$^{40}$ erg s$^{-1}$, which is similar to the kinetic power of the outflow (see Sect. 4.2). However, the typical SFRs of galaxies with candidate starburst-driven outflows are in the range  1 M$_{\odot}$ yr$^{-1}$ to hundred M$_{\odot}$ yr$^{-1}$ \citep[see e.g.][]{Cicone2014}. 
Attributing the outflow to the starburst thus requires a concentration of the galaxy-wide SFR in the inner 100~pc and/or an instantaneous SFR significantly larger than the long-term average SFR and a 100\% coupling of the supernova kinetic energy to the outflow. We thus conclude that starburst superwinds as the origin of the outflow in ESO 362-G18 are possible, but very unlikely.

\subsection{\textbf{Feeding versus feedback}}

  We can estimate the mass outflow rate as the ratio of the mass of the outflowing gas to the dynamical time at the nucleus, $M_{g}/t_{d}$. The gas mass is given by

\begin{equation}
    M_{g} = N_{e}m_{p}Vf,
\end{equation}

\noindent
where $N_{e}$ is the electron density, $m_{p}$ is the mass of the proton, $V$ is the volume of the region where the outflow is detected;  we fix this value as 0\farcs35 around the nucleus (i.e. within the seeing) and $f$ is the filling factor. The filling factor can be estimated from

\begin{equation}
L_{H\alpha} \approx f N_{e}^{2} j_{H\alpha}(T)V,
\end{equation}

\noindent
where $j_{H\alpha}$(T)=3.3534$\times$10$^{-25}$ erg cm$^{-3}$s$^{-1}$ \citep{Osterbrock1989} and $L_{H\alpha}$ is the H$\alpha$ luminosity emitted within the volume $V$. Substituting equation (2) into equation (1) we have

\begin{equation}
M_{g} = \frac{m_{p}L_{H\alpha}}{N_{e}j_{H\alpha}(T)}.
\end{equation}

In Section 4, we concluded that the high-velocity component detected both in [O\,{\sc iii}] and H${\alpha}$ are produced by the emission of clouds in front of the galactic disk, which is entrained by the AGN outflow. Therefore, we estimate the H${\alpha}$ luminosity from the highest velocity component, yielding $L_{\alpha}$ = 1.5 $\times$ 10$^{40}$ erg s$^{-1}$, considering a luminosity distance to ESO 362-G18 of 52.1 Mpc (from the NASA/IPAC Extragalactic Database). On the other hand, we assumed the $N_{e}$ as the mean value of 2453 cm$^{-3}$ within the inner 0\farcs35 from the nucleus. Taking into account all those values, we estimate an ionized gas mass of 1.5 $\times$ 10$^{4}$ M$_{\odot}$ for the outflow component.

The dynamical time $t_{d}$ can be estimated as the ratio of the radius where we are considering the outflow (0\farcs35 $\approx$ 86pc) to the mean deprojected velocity of the outflow ($\sim$394 km s$^{-1}$). This gives a $t_{d}$ of $\sim$ 2 $\times$ 10$^{5}$ years. Finally, the mass outflow rate $\dot{M}$ is 0.074 M$_{\odot}$ yr$^{-1}$. We point out that this $\dot{M}$ is only a lower limit since it corresponds only to the outflowing mass associated with the ionized side of the clouds. Additionally, if we consider a biconical outflow, assuming most of the far side of the bicone is hidden by the galactic disk, the $\dot{M}$ can be twice the calculated value, that is, $\dot{M}$ $\approx$ 0.15 M$_{\odot}$ yr$^{-1}$, in agreement with others $\dot{M}$ observed in nearby galaxies \citep{Barbosa2009, MullerSanchez2011,Lena2015}

We can now compare the mass outflow rate with the mass accretion rate required to feed the SMBH $\dot{M}$$_{acc}$, which can be estimated as

\begin{equation}
\dot{M}_{acc} = \frac{L_{bol}}{\eta c^{2}},
\end{equation}

\noindent
where $c$ is the speed of light and $L_{bol}$ is the bolometric luminosity. Using the $L_{bol}$ estimated by \citep{AgisGonzalez2014} of 1.3 $\times$ 10$^{44}$ erg s$^{-1}$, we derive a mass accretion rate of 2.2 $\times$ 10$^{-2}$ M$_{\odot}$ yr$^{-1}$, where we assume the radiative efficiency, $\eta$, to 0.1, the typical value derived from Shakura-Sunyaev accretion models onto a non-rotating black hole \citep{Shakura1973}. Therefore, the $\dot{M}$$_{acc}$ is $\sim$ 7 times lower than the mass outflow rate. This is consistent with previous works in nearby galaxies \citep{Barbosa2009,MullerSanchez2011} and indicates that most of the observed outflowing gas is mass entrained by the surrounding interstellar medium \citep{Veilleux2005,Storchi-Bergmann_review_2014}.

In a sample of 15 pairs of Seyfert plus inactive galaxies, \citet{Dumas2007} found that SMBHs with accretion rates larger than 10$^{-4.5}$ M$_{\odot}$ yr$^{-1}$ tend to lie in galaxies with  disturbed kinematics. However, in the case of ESO 362-G18, although the value of $\dot{M}$$_{acc}$ is almost a thousand times greater than 10$^{-4.5}$ M$_{\odot}$ yr$^{-1}$, large twists in the gas kinematics or significant misalignments between the gas and stellar kinematics are not observed. Thus, the nuclear activity in ESO 362-G18 may be related to major mergers in the past \citep{Hopkins_mergers2010} that do not leave current disturbances in its kinematics, rather than perturbations in the ionized gas or misalignments between the stellar and gas rotations \citep{Dumas2007}. The posited minor merger observed in the acquisition image (Fig. 1) produces no discernible disturbances in the gas or stellar kinematics within our nuclear FOV. 


Considering the galactic disk as well as the BLR from the broad component of Ha, the total ionized gas mass within $\sim$ 84 pc of the nucleus is $\sim$ 3.3 $\times$ 10$^{5}$ M$_{\odot}$. Assuming that this gas lies in the disk and that a small fraction ($\sim$ 10\%) suffers a radial inflow within the disk, infall velocities of $\sim$ 34 km s$^{-1}$  would be required to feed the outflow, the SMBH accretion, and maintain a SFR of 5.5 $\times$ 10$^{-3}$ M$_{\odot}$ yr$^{-1}$ \citep{Galbany2016} in ESO 362-G18. An inflow velocity of this magnitude (i.e. $\sim$20  km s$^{-1}$ in projection if it lies in the plane of the disk) is at the limit of detectability in our observations and analysis. 
The residual (observed - stellar model)  [N\,{\sc ii}] velocity map (rightmost panel of Fig. 6) shows blue (red) residual velocities on the far (near) side of the galaxy disk, which could be interpreted as a signature of inflow to the nucleus: the inflow velocities would then be of this order of magnitude. However, this residual pattern is primarily attributable to the mismatch between the PAs of the stellar and ionized gas kinematics (e.g. Fig. 1 of \citet{vanderKruit1978}); the residual 
(observed [N\,{\sc ii}] -  model [N\,{\sc ii}])  [N\,{\sc ii}] velocity map does not show this pattern.
One alternative to explain the relatively high accretion rate and the relatively low inflow rate is that the AGN is now passing through a period of maximum activity, which would be transient but cause an overestimation of the accretion rate and a greater required infall velocity to maintain it. Another possibility 
is that the total ionized gas mass is only the tracer of the true (dominated by molecular gas) amount of available gas. 

\subsection{\textbf{Kinetic power}}

Considering an outflow bicone, with a mass outflow rate ($\dot{M}$) of 0.148 M$_{\odot}$ yr$^{-1}$, we can obtain the kinetic power ($\dot{E}_{out}$) using the following expression

\begin{equation*}
\dot{E}_{out} = \frac{1}{2}\dot{M}(v^{2} + 3\sigma^{2}),
\end{equation*}

\noindent
where \textit{$v$} and \textit{$\sigma$} are the average velocity and velocity dispersion of the outflowing gas, respectively. Taking these values from the nuclear region ($\leqslant$ 0\farcs35) of the outflow component in H$\alpha$, we have \textit{v} = 394 km s$^{-1}$ and \textit{$\sigma$} = 74 km s$^{-1}$; then we obtain a kinetic power of $\dot{E}_{out}$ = 8 $\times$ 10$^{39}$ erg s$^{-1}$. 

With the aim to measure the effect (feedback) of the ionized gas outflow on the galactic bulge, we compare the kinetic power with the accretion luminosity ($L_{bol}$ = 1.3 $\times$ 10$^{44}$ erg s$^{-1}$),  obtaining a value of $\dot{E}_{out}/L_{bol}$ = 6.1 $\times$ 10$^{-5}$. This is at the lower end of the range (10$^{-4}$ -- 5 $\times$ 10$^{-2}$) found by \citet{MullerSanchez2011}\footnote{With the caveat that they used the following equation for the kinetic power: $\dot{E}_{out} = \frac{1}{2}\dot{M}(v_{max}^{2} + \sigma^{2})$ which, for ESO 362-G18, produces $\dot{E}_{out}/L_{bol}$ = 8.9 $\times$ 10$^{-5}$.}. Our lower ratio could be attributed to large uncertainties in the velocity dispersion ($\sigma$$_{inst}$ = 36 km s$^{-1}$) and the bolometric correction used to calculate L$_{bol}$ from $L_{X}$(2-10 keV) in \citet{AgisGonzalez2014}. This correction can vary widely (between 4 and 110) \citep[e.g.][Figs. 7 and 8]{Lusso2012}. 

\section{Summary and conclusions}

We observed the gaseous and stellar kinematics of the inner 0.7 $\times$ 1.2 kpc$^{2}$ of the nearby Seyfert 1.5 galaxy ESO 362-G18 using optical spectra (4092-7338 \AA) from the GMOS integral field spectrograph on the Gemini South telescope, which allows the detection of a number of prominent emission lines, i.e. H$\beta \lambda$4861, [O\,{\sc iii}]$\lambda \lambda$4959,5007, H$\alpha$+[N\,{\sc ii}] $\lambda \lambda$6548,6583 and [S\,{\sc ii}]$\lambda \lambda$6716,6731. We employed a variety of IDL and Python programmes to analyze these lines and obtain spatially resolved radial velocities, velocity dispersions, and fluxes at a spatial resolution of $\sim$170 pc and a spectral resolution of 36 km s$^{-1}$.  The main results of this paper are as follows.

\begin{itemize}
\renewcommand{\labelitemi}{$\bullet$}

 \item The H$\alpha$ and [O\,{\sc iii}] lines clearly show double-peaked emission lines near to the nucleus and to the SE. We used a two Gaussian fit to separate these profiles into two kinematic components: a low-velocity component and high-velocity component.
 
 \item The stars, [N\,{\sc ii}] and [S\,{\sc ii}] emission lines, and low-velocity component of H$\alpha$ and [O\,{\sc iii}] lines typically have radial velocities between -80 km s$^{-1}$ and 70 km s$^{-1}$, and have very similar rotation patterns, so we interpret all of these to originate in the rotating galactic disk. 
 
 \item The high-velocity component of H$\alpha$ and [O\,{\sc iii}] reach values in excess of 200 km s$^{-1}$ with respect to the systemic velocity, and we argue that these spectral components originate from gas outflowing within the AGN radiation cone. We present a toy model to explain why this gas is preferentially redshifted to our LOS, except at the nucleus where blueshifted [O\,{\sc iii}] emission from the outflow traces the region where the outflow is still breaking out of the galactic disk. The effects of AGN ionization has been previously observed, showing a fan-shaped morphology with an extension of $\approx$ 10\arcsec\ to the SE in emission-line and excitation maps.
  
 
  \item The assumption that the outflow component is behind the plane of the sky is also motivated by the velocity dispersions observed in [O\,{\sc iii}]: while the disk component presents the highest dispersions in most of our FOV, the outflow component exceeds it in the nuclear region, where the highest blueshift velocities are reached. This difference is consistent with attenuation from the galactic disk except very close to the nucleus, where the approaching side of the cone can be seen. 
  
 \item The structure of the nuclear region of ESO 362-G18 presents spiral arms in a trailing pattern, which are increasingly fainter as we approach the nucleus. Considering the linear scale of our observations, we posit that the unusual dust morphology is a result of instabilities produced near to the inner Lindblad resonance, which is expected within the inner 1.6kpc. The presence of the dust structures also supports the hypothesis of \citep{Simoes2007} that the presence of dust is a necessary condition for the nuclear activity in AGNs.

\item While morphologically there is evidence that ESO 362-G18 is participating in a minor merger, we do not find any effect of this in the stellar or gas kinematics within our relatively small FOV. 
 
\item Using the H${\alpha}$ luminosity, we estimate a lower limit for the mass outflow rate $\dot{M}$ of 0.074 M$_{\odot}$ yr$^{-1}$. This value will double if a biconical outflow is assumed. Further, the value we calculate is likely a lower limit to the outflow mass and rate, as we have argued that our H$_\alpha$ luminosity used to calculate the outflow gas mass only represents the fraction of the NLR gas clouds illuminated by the AGN, rather than all of the outflowing NLR gas. In any case, our estimated outflow rate is significantly higher than the accretion rate necessary to sustain the AGN bolometric luminosity, that is, $\dot{M}$$_{acc}$ $\sim$ 2.2 $\times$ 10$^{-2}$ M$_{\odot}$ yr$^{-1}$.

  \end{itemize}
   
\begin{acknowledgements}
  This work is based on observations obtained at the Gemini Observatory, which is operated by the Association of Universities for Research in Astronomy, Inc., under a cooperative agreement with the NSF on behalf of the Gemini partnership: the National Science Foundation (United States), the Science and Technology Facilities Council (United Kingdom), the National Research Council (Canada), CONICYT (Chile), the Australian Research Council (Australia), Ministerio da Ci\^{e}ncia e Tecnologia
(Brazil) and south-eastCYT (Argentina). NN gratefully acknowledges support from the Chilean BASAL Centro de Excelencia en Astrofísica y Tecnologías Afines (CATA) grant PFB-06/2007. PH, NN, PS and DM acknowledge support from Fondecyt 1171506. VF acknowledges support from CONICYT Astronomy Program-2015 Research Fellow GEMINI-CONICYT (32RF0002). 
R.A.R. acknowledges support from FAPERGS (project No. 16/2551-0000251-7) and CNPq (project N0. 303373/2016-4). 
This research has made use of the NASA/IPAC Extragalactic Database (NED) which is operated by the Jet Propulsion Laboratory, California Institute of Technology, under contract with the National Aeronautics and Space Administration.
\end{acknowledgements}

\bibliographystyle{aa} 
\bibliography{aanda} 

\end{document}